\documentclass{statsoc} 
\RequirePackage{amsmath,amssymb} 
\RequirePackage{txfonts}
\RequirePackage[T1]{fontenc}
\RequirePackage{mathtools}
\RequirePackage[authoryear]{natbib}
\RequirePackage[colorlinks,citecolor=blue,urlcolor=blue]{hyperref}
\RequirePackage{bm}
\RequirePackage{graphicx}
\DeclareGraphicsExtensions{.pdf}

\newcommand\ie{i.\,e.}

\newcommand\eg{e.\,g.}
\newcommand\Eg{E.\,g.}
\newcommand\motabar{MOTABAR}

\newcommand\fmath[1]{\fbox{$\displaystyle #1$}}

\newcommand{\ui}[1]{^{\text{\guilsinglleft}#1\text{\guilsinglright}}}
\newcommand\reals{\mathbb{R}}
\newcommand\naturals{\mathbb{N}}
\renewcommand\ge{\geqslant}
\renewcommand\geq{\geqslant}
\renewcommand\le{\leqslant}
\renewcommand\leq{\leqslant}
\newcommand\eps{\varepsilon}
\renewcommand\rho{\varrho}
\renewcommand\phi{\varphi}

\newcommand\diag{{\rm diag}}
\newcommand\sgn{{\rm sign}}

\newcommand\Cov{{\rm Cov}}
\newcommand\Var{{\rm Var}}
\newcommand\E{{\rm E}}
\newcommand\prob{\mathbb{P}}
\newcommand\erf{\text{erf}}
\newcommand\wang{{\rm Wang}}
\newcommand\corr{{\rm corr.}}
\newcommand\trace{{\rm trace}}

\newcommand{\python}[1]{}

\title[Moving Taylor Bayesian Regression]{Moving Taylor Bayesian Regression for nonparametric multidimensional function estimation with possibly correlated errors}

\author[Jobst Heitzig]{Jobst Heitzig}
\runauthor{Jobst Heitzig}
\address{Potsdam Institute for Climate Impact Research (PIK),
Transdisciplinary Concepts and Methods,
P.O.\,Box 60 12 03, 14412 Potsdam, Germany.}
\email{heitzig@pik-potsdam.de}

\begin{document}

\begin{abstract}
	We present a nonparametric method for estimating the value and several derivatives of 
    an unknown, sufficiently smooth real-valued function of real-valued arguments 
    from a finite sample of points,
    where both the function arguments and the corresponding values 
    are known only up to measurement errors having some assumed distribution and correlation structure.
    The method, \textit{Moving Taylor Bayesian Regression} (\motabar), 
    uses Bayesian updating to find 
    the posterior mean of the coefficients of a Taylor polynomial of the function 
    at a moving position of interest.
  	When measurement errors are neglected, \motabar\ becomes a multivariate interpolation method.
  	It contains several well-known regression and interpolation methods as special or limit cases.
  	We demonstrate the performance of \motabar\ 
  	using the reconstruction of the Lorenz attractor from noisy observations as an example.
\end{abstract}
\keywords{interpolation; irregular sampling; numerical differentiation; rational function; smoothing}



\section{Introduction}

\subsection{Motivation}

The basic task of regression analysis 
-- the estimation of values of an unknown, sufficiently smooth function $f$ of one or several real arguments,
given a finite amount of possibly noisy data -- 
occurs pervasively in many kinds of quantitative scientific research.
Most existing approaches to this task can roughly be classified into two groups:
{\em parametric} or model-based approaches such as polynomial regression, spline smoothing \citep{Reinsch1967a}, 
or kriging \citep{Krige1951},
and {\em nonparametric} approaches such as local regression, nearest or 
natural neighbour estimation \citep{Sibson1981}, 
inverse distance weighting \citep{Shepard1968}, 
or kernel smoothing. 
Although parametric methods are usually based on more rigorous reasoning 
such as maximum likelihood estimation, Bayesian updating, approximation theory, 
or some other kind of optimization,
their results are only guaranteed to be reliable if the sought function is assumed to belong to
some particular model or function class with a small number of parameters,
\eg, polynomials or splines of a fixed degree and fixed set or number of break points.
If, as is often the case in empirical research, this strong assumption can not be justified,
nonparametric methods are available which, however, 
are usually based on various forms of more heuristic reasoning, 
frequently involve the choice of several control parameters such as 
the number of neighbours or the choice of a weight or kernel function,
and sometimes do not provide an easily interpreted assessment of reliability such as 
a standard error or confidence interval.
Also, many regression methods are not or only restrictedly applicable 
if one or several of the following conditions apply: 
(i) there might be measurement errors not only in the function values (the ``dependent'' variable) 
but also in the function arguments (the ``independent'' variables),
(ii) measurement errors might not be independent and identically distributed,
(iii) the function is of more than one real argument,
(iv) the measured sample is irregularly distributed in the function's argument space,  
and/or (v) also several derivatives of the function shall be estimated.  

\begin{figure}\begin{centering}%
\includegraphics[width=0.49\textwidth]{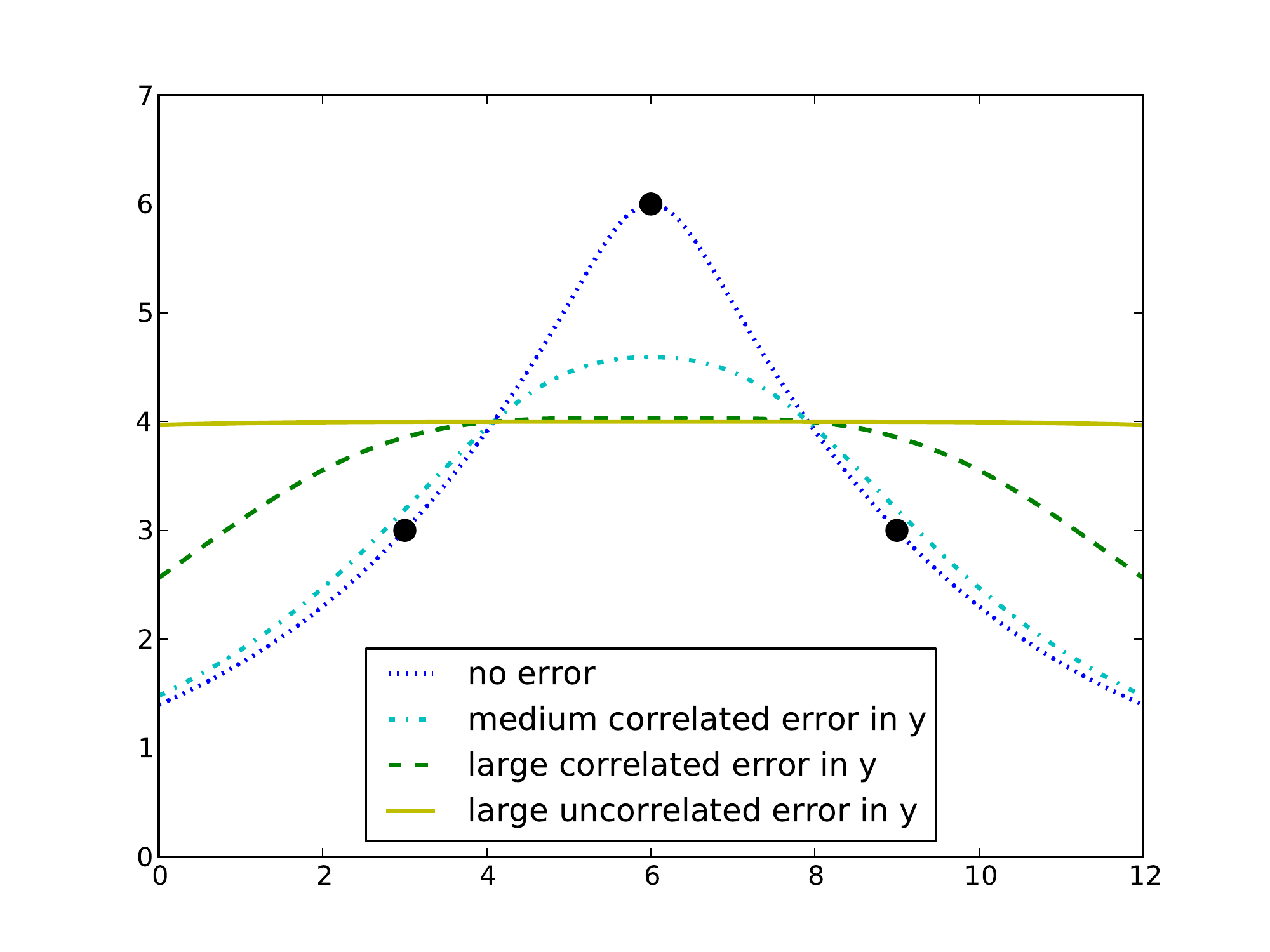}%
\includegraphics[width=0.49\textwidth]{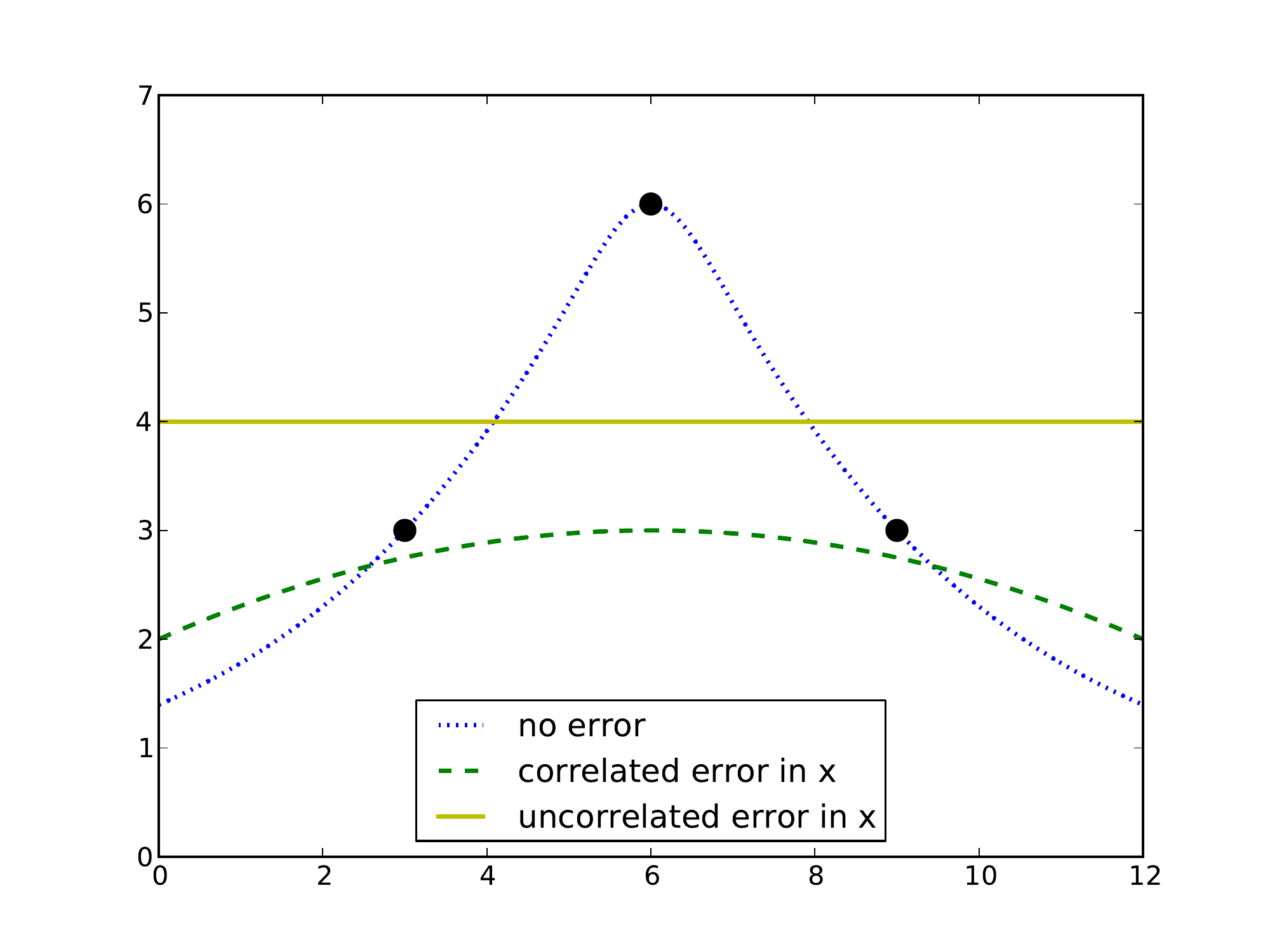}%
\end{centering}
\caption{\label{fig:hat3}
(Colour online) 
Illustration of the effect of error correlation on some \motabar\ estimates.
}\end{figure}

In order to illustrate how plausible estimates can depend on 
the assumed amount and correlation of measurement errors, 
consider the minimal data $(3,3)$, $(6,6)$, and $(9,3)$
and assume that we want to estimate $f$ and $f'$ on the interval $[0,12]$ on the basis of this data.
In the left diagram in Fig.\,\ref{fig:hat3}, four different such estimates of $f$ are shown,
which were produced with the method we will describe in this article,
using different assumptions on the measurement error contained in the data.
Without measurement errors, \ie, if both $x$ and $y$ data are precise, 
it is plausible to estimate $f(3)=3$ and $f'(3)\approx 1$, as on the blue dotted line.
Without errors in $x$ but with large independent errors in $y$,
the apparent slope becomes quite uncertain,
and one would rather estimate $f(3)$ by the sample mean, $4$, as on the yellow line.  
However, if the errors in $y$ are highly correlated,
\eg, because of a systematic but unknown bias in the measurement equipment, 
then one knows that the data is basically only shifted in the $y$ direction,
which has no influence on slopes,
so one would probably expect $f'(3)>0$ again, as on the green dashed line.
The smaller the errors, the more the slope should resemble $1$, as on the cyan dash-dotted line. 
Similarly, if the $y$ data is precise but the $x$ data is highly uncertain,
one would use the sample estimate if errors are uncorrelated,
as on the yellow line in the right diagram in Fig.\,\ref{fig:hat3}.
If errors in $x$ are correlated, one would retain some slope information
which suggests that the values to the left and right of the three data points 
are rather below than above three.
Hence the estimate would be lower than with uncorrelated errors,
as on the green dashed line.
The occurrence of this lowering effect 
also shows that the effect of errors in the $x$ dimension is usually not the same as 
the effect of some ``equivalent'' amount of errors in the $y$ dimension.

\subsection{Moving Taylor Bayesian Regression}

In this article, we develop a regression method which 
is nonparametric in the sense that it can estimate any sufficiently smooth function,
but is still based on rigorous statistical reasoning,  
can deal with any of the above situations (i)--(v),
and contains several existing methods as special or limiting cases.
The method, {\em Moving Taylor Bayesian Regression (\motabar)}, is based on the following ideas:
\begin{enumerate}
  \item Approximate $f$ locally by a Taylor polynomial at a moving position of interest $\xi$.
  \item Treat the unknown Taylor coefficients as parameters in a statistical model.
  \item Use the measured data to update prior beliefs about these parameters using Bayesian updating.
  \item Use the posterior mean (or mode) and variance of the parameters 
        as estimates of the function's value and derivatives and the corresponding estimation uncertainty.
\end{enumerate}
In principle, these steps can be performed for any form of error and prior distribution,
but the method becomes particularly transparent 
if the value measurement error is assumed to be multivariate Gaussian
and the involved prior distributions are also multivariate Gaussian.
In that case, the resulting posterior distributions can be written as mixtures of Gaussians
whose mean and variance can be derived analytically,
and the resulting \motabar\ estimator $\hat f(\xi)$ of $f(\xi)$
turns out to be a mixture of smooth rational functions of $\xi$
that can be computed using simple linear algebra.
The mixing is related to the distribution of argument measurement errors.
Without argument measurement errors, $\hat f(\xi)$ is a smooth rational function of $\xi$
whose algebraic form can be seen as a generalization of 
ordinary least squares regression estimators.
For these reasons, we will focus on the case of Gaussian priors and value errors in this article.

The input data needed to apply \motabar\ are then
\begin{itemize}
  \item the measured data,
  \item variances and possibly covariances of both argument and value measurement errors,
  \item a prior covariance matrix for the value of $f$ and its derivatives of order $<p$ for some $p>0$,
  \item prior variances for $f$'s derivatives of order $p$, 
  \item and, as the only free control parameter, an integer $p$ 
  	    larger than the order of any derivative of $f$ one wants to estimate.
\end{itemize}

For particular cases of error and prior distributions, 
it will turn out that the \motabar\ estimate equal or approximate 
the results of some well-known other methods, 
including ordinary polynomial regression, inverse distance weighting, and linear interpolation on regular grids.
This behaviour of our estimator is similar to that of a related approach by \citet{Wang2010c}
in which $f$ is also approximated by a Taylor polynomial at a moving position of interest $\xi$,
but in which the Taylor coefficients are then estimated not by Bayesian updating 
but by minimizing a heuristic loss function motivated by approximation theory
(see Sec.\,\ref{sec:wang}).
Other cases of error and prior distributions result in plausible generalizations or variants of well-known methods,
\eg, a new form of local polynomial interpolation and a Bayesian variant of inverse distance weighted smoothing.

In practise, the needed prior and error variances and covariances 
might themselves be estimated from other or even the same data,
a question we do however not address in detail in this article.


\paragraph{Comparison with other methods.}

\motabar\ can be interpreted as a kind of {\em local} regression since
although it takes into account also data points far away from $\xi$,
it gives them much less influence on the estimate than those close to $\xi$.
This fact is reflected in the occurrence of a {\em weight matrix} $W$ in the estimator equation.
But unlike other local or piece-wise methods such as nearest or natural neighbour regression, 
locally weighted scatterplot smoothing (LOESS) \citep{Cleveland1979}, or splines,
the \motabar\ estimate $\hat f(\xi)$ is infinitely smooth (\ie, infinitely often differentiable),
at least as long as the value measurement errors have nonzero variance, 
and there are either no argument measurement errors 
or their probability density is infinitely smooth 
(\eg, when argument measurement errors are Gaussian as well). 
This is because in \motabar\, 
the weight of each individual data point in the estimate depends smoothly on its distance from $\xi$,
whereas in other methods the weights can switch from zero to nonzero in a nonsmooth way as $\xi$ moves.

The nonsmooth change in weights that other methods involve is also counter-intuitive 
when arguments can only be measured with some error.
\Eg, suppose that measurements resulted in the argument-value pairs 
$(-1,0)$, $(0,1)$, $(1,0)$, and $(1/100,-1)$,
where the argument measurements involved an error of magnitude $1/10$,
so that the last measurement might actually reflect $f(-1/100)$ instead of $f(1/100)$. 
A linear interpolant of the four measurements would have a slope of $\approx +1$,
but when the $(1/100,-1)$ measurement is moved towards $(1/100,-1)$,
the slope would discontinuously switch to $\approx -1$.
A similar effect occurs with splines.
In other words, when the ranking of the argument measurements with respect to their distance from $\xi$ is uncertain 
due to measurement errors,
it seems inappropriate to use a method that strongly relies on the correctness of this ranking,
\eg, by using only the $k$ nearest neighbours of $\xi$;
although disregarding faraway measurements or extreme observations completely 
instead of just downweighting them
might still be advisable to increase the robustness of the method
if outliers might exist, \eg, due to fat-tailed error distributions.

One existing class of methods, Inverse Distance Weighting (IDW) \citep{Shepard1968}, 
also uses smoothly decaying weights that only depend on the distance from $\xi$.
In a commonly used variant of IDW, 
the weights are inversely proportional to the square or a larger power of the distance,
and we will show that this method can be derived as a special case of \motabar\ 
in which a noninformative prior distribution for $\phi$ is used.

A common feature of many interpolation and smoothing methods, 
including IDW and other weightings- or kernel-based methods, linear interpolation, and nearest or natural neighbour interpolation, 
is that the estimate cannot exceed the largest measured values, 
even if the data strongly suggest a nonzero slope at the largest data point.
Such methods will therefore always underestimate the maxima of $f$.
Other methods, like polynomial regression, 
can in some situations `overshoot' and result in estimates that lie far outside the measured range.
Spline methods are often considered a good compromise between these two behaviours regarding maxima. 
Depending on the choice of $p$, \motabar\ will behave quite similar to spline methods in this respect.
As a consequence, the weights of individual data values in the estimate might be $<0$ or $>1$.

Similar to other nonparametric methods or methods with many parameters, 
the \motabar\ estimate might fit the data too narrowly and show too much fluctuations due to this `overfitting'
when $p$ and the prior variances are badly chosen.
If the prior variances for higher derivatives grow too fast, 
the estimate is allowed to vary on smaller scales than the sampling density can resolve reliably in view of the assumed error distributions.
The amount of overfitting can thus be controlled by varying $p$ and the priors.
%




\subsection{Framework}

Assume that we are interested in the value of a certain function $f:\reals^d\to\reals$
and all its derivatives of order $<p$ at a certain {\em position of interest} $\xi\in\reals^d$,
where $p>0$ and $f$ is assumed to be $p$ times continuously differentiable.
We take the convention to write elements of $\reals^d$ as column vectors
and to enumerate their $d$ components with a subscript index,
hence $\xi=(\xi_1,\dots,\xi_d)'$ where the $'$ symbol denotes transposition.
Assume that the only information we have about $f$ is 
(i) $N$ {\em measured data points} $(x\ui i,y\ui i)$ with $i=1\ldots N$,
(ii) some estimates of the magnitude of {\em measurement errors}, and
(iii) some beliefs about the variability of $f$ and some of its derivatives.
These assumptions will be made more precise later.
We enumerate measurements with superscript indices in guillemots, 
hence $x\ui i=(x\ui{i}_1,\dots,x\ui{i}_d)'$.
Assume the $i$-th data point $(x\ui i,y\ui i)$ 
is the result of trying to measure $f$ at the argument $x\ui i$,
but the measurement might involve an error in both the argument and the value,
so that the actual result $y\ui i$ is the value of $f$ 
at a slighly different argument $\chi\ui i=x\ui i-\gamma\ui i$ (where $\gamma\ui i\in\reals^d$ is the {\em argument error})
plus some {\em value error} $\eps\ui i\in\reals$.
In other words,
\begin{align}
	y\ui i &= f(x\ui i-\gamma\ui i)+\eps\ui i.
\end{align}

\paragraph{Multi-index notation.}
For dealing with higher-order derivatives and multidimensional Taylor polynomials of $f$, 
it is convenient to use a {\em multi-index}
$\alpha\in\naturals^d$ consisting of nonnegative integers $\alpha_k\geq 0$ for $=1\dots d$.
We then use the denotations
\begin{align}
	\alpha &= (\alpha_1,\alpha_2,\dots,\alpha_d)', & 
	|\alpha| &= \alpha_1+\alpha_2+\cdots+\alpha_d\in\naturals,\\
	\alpha! &= \alpha_1!\alpha_2!\cdots\alpha_d!\in\naturals, & 
	x^\alpha &= x_1^{\alpha_1}x_2^{\alpha_2}\cdots x_d^{\alpha_d}\in\reals, 
\end{align}
\begin{align}
	(D^\alpha f)(x) &= \frac{\partial^{\alpha_1}}{\partial x_1^{\alpha_1}}\frac{\partial^{\alpha_2}}{\partial x_2^{\alpha_2}}\cdots
						\frac{\partial^{\alpha_d}}{\partial x_d^{\alpha_d}}f(x_1,\dots,x_d)\in\reals,
\end{align}
where $x\in\reals^d$ and $|\alpha|\le p$.
Note that the order of differentiation in $(D^\alpha f)(x)$ is unimportant for $|\alpha|\le p$.
Our quantities of interest are then the derivatives
\begin{align}
	\phi\ui{\alpha} &= (D^\alpha f)(\xi) & (|\alpha|<p).\label{eqn:phi}
\end{align}

\section{Using Taylor's Theorem to get a local model of the function at the position of interest}

For a given position of interest $\xi\in\reals^d$,
a relationship between our quantities of interest $\phi\ui\alpha$ 
and the data $x\ui i,y\ui i$ is given by Taylor's Theorem, which leads to 
\begin{align}\label{eqn:taylor}
	y\ui{i} &= \sum_{\alpha,|\alpha|<p}X\ui{i,\alpha}\phi\ui{\alpha}
			+ r\ui{i}
			+ \eps\ui{i},
\end{align}
where 
\begin{align}
	X\ui{i,\alpha} &= \frac{(\chi\ui{i}-\xi)^\alpha}{\alpha!}, &
    r\ui{i} &= \sum_{\alpha,|\alpha|=p}X\ui{i,\alpha}\psi\ui{\alpha i},\label{eqn:r}\\
	\psi\ui{\alpha i} &= (D^\alpha f)(\xi+\lambda\ui{\alpha i}(\chi\ui{i}-\xi)), &
	\lambda\ui{\alpha i} &\in[0,1]\quad (|\alpha|=p).
\end{align}
Note that there are $m={p-1+d\choose d}$ many choices for $\alpha$ with $|\alpha|<p$
and $p-1+d\choose d-1$ many choices for $\alpha$ with $|\alpha|=p$.
It will be convenient to stack all relevant quantities into the column vectors
\begin{align}
	\bm x &=(x\ui{1}_1,\dots,x\ui{1}_d,\dots,x\ui{N}_1,\dots,x\ui{N}_d)',\\
	\bm\gamma &=(\gamma\ui{1}_1,\dots,\gamma\ui{1}_d,\dots,\gamma\ui{N}_1,\dots,\gamma\ui{N}_d)',
\end{align}
$\bm\chi = \bm x-\bm\gamma$,
$\bm y =(y\ui{i},\dots,y\ui{N})'$, 
$\bm\eps =(\eps\ui{1},\dots,\eps\ui{N})'$,
$\bm r =(r\ui{1},\dots,r\ui{N})'$,
\begin{align}
	\bm\phi &=(\phi\ui{\alpha}:|\alpha|<p)', & 
    \bm\psi &=(\psi\ui{\alpha i}:|\alpha|=p,i=1\dots N)',
\end{align}
and the matrix
\begin{align}
    X &= (X\ui{i,\alpha}:i=1\dots N,|\alpha|<p)\label{eqn:X}.
\end{align}
Note that the column vector $\bm\psi$ has a row for each combination of $\alpha$ and $i$, 
and we write this row index as $\alpha i$. 
This must not be confused with the row-and-column index pair $i,\alpha$ of the matrix $X$.
In other words, $\bm\phi$ is an $m\times 1$ vector,
$\bm\psi$ is ${p-1+d\choose d-1}N\times 1$,
and $X$ is an $N\times m$ matrix.
Eq.\,\ref{eqn:taylor} is now summarised in matrix notation as
\begin{align}\label{eqn:taylorshort}
	\bm y &= X \bm\phi + (\bm r + \bm\eps).
\end{align}
Although this is formally a linear regression model,
we are not interested in estimating its coefficient matrix $X$ 
(which we know already up to some measurement errors), 
but in estimating the regressor $\bm\phi$,
and this we need to do for each position of interest $\xi$ separately.
This is the reason why we next apply Bayes' Theorem to Eq.\,\ref{eqn:taylorshort}.

\section{Using Bayesian updating to estimate the function value and derivatives}

\subsection{General approach}

Let us model our information about the quantities of interest $\bm\phi\ui\alpha = (D^\alpha f)(\xi)$
and about the other unknown terms in Eq.\,\ref{eqn:taylorshort}
as Bayesian beliefs, \ie, in the form of (subjective) probability distributions,
by treating $\bm\phi$ and $\bm\psi$ as random variables with Lebesgue-integrable densities $\rho(\cdot)$.
Because in general, each $\psi\ui{\alpha i}$ corresponds to an argument $\xi+\lambda\ui{\alpha i}$ that is different from $\xi$, 
we assume $\bm\phi$ and $\bm\psi$ are independent.
Also, we assume that the function-related quantities $\bm\phi$ 
are independent from the measurement-related quantities $\bm x$ and $\bm\gamma$.  
The measured data $\bm x,\bm y$ then allow us to update our prior beliefs $\rho(\bm\phi)$ about $\bm\phi$
by applying Bayes' Theorem to Eq.\,\ref{eqn:taylorshort},
which leads to posterior beliefs 
\begin{align}\label{eqn:bayes}\fmath{
    \rho(\bm\phi|\bm x,\bm y) \propto 
            \rho(\bm\phi)
            \int_{\reals^{Nd}} d^{Nd}\bm\gamma\,\rho(\bm\gamma|\bm x)
            \int_{\reals^N} d^N\bm r\,\rho(\bm r|\bm x,\bm\gamma)\rho(\bm y|\bm x,\bm\gamma,\bm\phi,\bm r).
}\end{align}
Note that we follow the general convention to use the same symbol, here $\rho$, to refer to all occurring probability densities
since it will always be clear from the context which variable's density is meant.
Also, we suppress the domain of integration in the following,
and we are not interested in multiplicative constants that do not depend on $\bm\phi$,
and use the symbol $\propto$ to denote equality up to such a constant.
To utilise Eq.\,\ref{eqn:bayes}, we need to specify  
\begin{itemize}
\item a distribution $\rho(\bm\gamma|\bm x)$ for the argument errors that might depend
on the arguments,
\item a distribution $\rho(\bm\eps|\bm x,\bm\gamma)$ of the value errors that might depend
on the arguments and argument errors, and
\item prior distributions $\rho(\bm\phi)$ and $\rho(\bm\psi)$ 
expressing our initial information about the values and derivatives of $f$ at $\xi$ before the measurements.
\end{itemize}
The term $\rho(\bm r|\bm x,\bm\gamma)$ can then be determined from $\rho(\bm\psi)$ using Eq.\,\ref{eqn:r},
while the term $\rho(\bm y|\bm x,\bm\gamma,\bm\phi,\bm r)$ can be determined from $\rho(\bm\eps|\bm x,\bm\gamma)$ using Eq.\,\ref{eqn:taylorshort}.
As an estimate of $\bm\phi$ one can then use any measure of central tendency of the posterior distribution as given by Eq.\,\ref{eqn:bayes},
\eg, the posterior mean, median, or mode,
while an estimate of the estimation error would be given by a suitable measure of dispersion, 
\eg, the posterior variance or the median distance from the median.
Although this strategy can in principle be applied to error and prior distributions of any form,
our approach becomes especially simple in the Gaussian case.

\subsection{Gaussian value error and priors}

For the rest of this article, we will assume that value errors and priors are 
(multivariate) Gaussian,
\begin{align}\label{eqn:gaussian}
    \rho(\bm\eps|\bm x,\bm\gamma) &\propto \exp(-\bm\eps' P_{\eps}\bm\eps/2),\\
    \rho(\bm\phi) &\propto \exp\{-(\bm\phi-\bm\mu_\phi)' P_\phi (\bm\phi-\bm\mu_\phi)/2\},\label{eqn:phiprior}\\
    \rho(\bm\psi) 
       &\propto \exp\{-(\bm \psi-\bm\mu_\psi)' P_\psi (\bm \psi-\bm\mu_\psi)/2\},
\end{align} 
where we use {\em precision matrices} $P_{\cdot}=\Sigma_\cdot^{-1}$ 
and omit to denote the dependency on $\bm x,\bm\gamma$ to simplify the following equations. 
Note that $P_\eps$ is a $Nd\times Nd$ square matrix whose rows and columns we address using 
indices of the form $ik$ with $i\le N$ and $k\le d$.
The case in which it is known that there are no value errors requires some
special treatment since then $\Sigma_\eps=0$ so that $P_\eps$ does not exist.
In that case, $\rho(\bm\eps|\bm x,\bm\gamma)$ behaves like a Dirac delta function,
giving $\int d^N\bm\eps\,\rho(\bm\eps|\bm x,\bm\gamma)g(\eps)=g(0)$
for all integrable functions $g$ of $\bm\eps$.

The Taylor remainders $\bm r$ given $\bm x$ and $\bm\gamma$
are then also Gaussian with mean $\bm\mu_r$ and covariance matrix $\Sigma_r$ given by 
\begin{equation}\label{eqn:postr}
    \mu_r\ui{i} = \underset{|\alpha|=p}{\sum_\alpha} X\ui{i,\alpha}\mu_\psi\ui{i,\alpha},\quad
    \Sigma_r\ui{i,j} = \underset{|\alpha|=|\beta|=p}{\sum_\alpha\sum_\beta}
        X\ui{i,\alpha}\Sigma_\psi\ui{\alpha i,\beta j}X\ui{j,\beta}.
\end{equation} 
Note that for almost all choices of $\bm\gamma$ and $\Sigma_\psi$, the matrix $\Sigma_r$ is nonsingular.
We will therefore assume that $\bm\gamma$ has a continuous distribution and $\Sigma_\psi$ is nonsingular, 
so that almost surely $\Sigma_r$ is nonsingular.
If it is known that $\bm\gamma= 0$, we assume instead that $\bm x,\xi$ are in general position,
in which case $\Sigma_r$ is also nonsingular.

To get an understanding of the relative sizes of the entries in $\Sigma_r$,
consider the special case in which $\Sigma_\psi$ is block-diagonal with identical $N\times N$ blocks $V$, 
so that $\Sigma_\psi\ui{\alpha i,\beta j}=\delta_{\alpha\beta}V\ui{i,j}$. 
Then $\Sigma_r\ui{i,j} = V\ui{i,j}\{(\chi\ui{i}-\xi)'(\chi\ui{j}-\xi)\}^p/p!\neq 0$,
showing how the covariance of the remainders $r\ui{i}$ and $r\ui{j}$ grows with the $p$-th 
power of the scalar product of $\chi\ui{i}$ and $\chi\ui{j}$. 
If $V$ is diagonal, also $\Sigma_r$ is diagonal 
and the variance of the remainder $r\ui{i}$ grows with the $2p$-th power of 
the distance between the actual argument of measurement $\chi\ui{i}$ and the position of interest $\xi$.

Now the main step to solving Eq.\,\ref{eqn:bayes} is to see that 
the posterior of $\bm\phi$ given $\bm x$, $\bm y$, and $\bm\gamma$ is
still Gaussian,
\begin{align}
    \rho(\bm\phi|\bm x,\bm y,\bm\gamma) &\propto
        \exp\{-(\bm\phi-\bm{\tilde\mu}_{\phi})' \tilde P_{\phi} (\bm\phi-\bm{\tilde\mu}_{\phi})/2\},
\end{align}
with precision matrix $\tilde P_\phi = \tilde P_{\phi}(\bm\gamma)$ 
and mean $\bm{\tilde\mu}_{\phi} = \bm{\tilde\mu}_{\phi}(\bm\gamma)$ given by
\begin{align}\label{eqn:givengamma}
    \fmath{
    \tilde P_{\phi}(\bm\gamma)  = P_\phi + X' W X,\quad
    \bm{\tilde\mu}_{\phi}(\bm\gamma) =
       \tilde P_{\phi}^{-1}\left\{P_\phi\bm\mu_\phi + X'W(\bm y - \bm\mu_r)\right\},
    }
\end{align}
where we call 
\begin{equation}\label{eqn:w}
	W=(\Sigma_r+\Sigma_\eps)^{-1}
\end{equation}
the {\em squared weight matrix}.
Note that the basic behaviour of $W\ui{i,j}$ is to decrease with 
a growing distance of $\chi\ui{i}$ and $\chi\ui{j}$ from $\xi$,
similar to the weights used in IDW.

\paragraph{Singularities.}
Although in pathological cases, some of the involved matrices can be singular,
they are nonsingular in general.
More precisely, assume that
$\bm\gamma$ has a continuous distribution and $\Sigma_\psi$ is nonsingular,
so that $\Sigma_r$ is nonsingular, too (see Eq.\,\ref{eqn:postr}).
Then $W$ exist and is nonsingular for almost all choices of $\Sigma_\eps$,
including the case $\Sigma_\eps=0$.
In that case, and for generic $\bm\chi$ and $\xi$, 
$X$ has full rank $\min(N,m)$ (like a Vandermonde matrix), 
so that also $X' W X$ is nonsingular when $m\le N$.
Finally, under these assumptions 
$\tilde P_{\phi}$ is nonsingular for almost all choices of $P_\phi$,
including the case $P_\phi=0$.

\subsection{The Moving Taylor Bayesian Regression estimator}

Integrating over all possible values of the argument errors $\bm\gamma$ finally shows 
that the posterior of $\bm\phi$ given $\bm x$ and $\bm y$ is a mixture of Gaussians
whose mean and covariance matrix are
\begin{align}
    &\fmath{\bm{\hat\phi} =  
        \int d^N\bm\gamma\,\rho(\bm\gamma|\bm x)
        \bm{\tilde\mu}_\phi(\bm\gamma),}\label{eqn:estimator}\\
    &\fmath{\hat\Sigma_\phi 
    =  \int d^N\bm\gamma\,\rho(\bm\gamma|\bm x)\left\{
            \tilde P_\phi(\bm\gamma)^{-1} + 
            (\bm{\tilde\mu}_\phi(\bm\gamma)-\bm{\hat\phi})
            (\bm{\tilde\mu}_\phi(\bm\gamma)-\bm{\hat\phi})'
        \right\}.}\label{eqn:sigmapost}
\end{align} 
This posterior mean, which we call the {\em \motabar\ estimator},
can now be used as a natural estimate of $\bm\phi$ given $\bm x$, $\bm y$,
and one can see that this estimate is 
\begin{itemize}
  \item an affinely linear function of the measured values $y\ui{i}$,
        with weights that decrease with growing distance between $\xi$ and $x\ui i$, and
  \item a mixture of values $\bm{\tilde\mu}_{\phi}(\bm\gamma)$ 
    each of which is an infinitely smooth rational function of the position of interest $\xi$
    that has no poles.
\end{itemize}
Because the entries of the inverse of an $m\times m$ square matrix $A$
are rational functions of the entries of $A$ of degree at most $(m-1,m)$,
the degree of $\bm{\tilde\mu}_{\phi}(\bm\gamma)$ is at most $(2p^2N,2p^2N)$. 
In practice, one can determine $\bm{\tilde\mu}_{\phi}(\bm\gamma)$ and $\tilde P_\phi(\bm\gamma)$ like this,
avoiding the large $N\times N$ matrix inversions and solving instead two systems of linear equations:
\begin{enumerate}
  \item Given $\xi$, $\bm x$, and $\bm\gamma$, compute $X$, $\bm\mu_r$, and $\Sigma_r$ from 
    Eqns.\,\ref{eqn:X} and \ref{eqn:r}.
  \item Determine $A=WX$ and $B=W(\bm y - \bm\mu_r)$ by solving 
        $(\Sigma_r + \Sigma_\eps)(A,B) = (X,\bm y - \bm\mu_r)$.
  \item Compute $\tilde P_{\phi} = P_\phi + X'A$.
  \item Determine $\bm{\tilde\mu}_{\phi}$ by solving 
    $\tilde P_{\phi}\bm{\tilde\mu}_{\phi} = P_\phi\bm\mu_\phi + X'B$.
\end{enumerate}
Note that in pathological cases, either of the two systems might not be solvable uniquely,
which possibility we do not discuss here.
Although the usually smaller $m\times m$ matrix inversions needed to determine $\hat\Sigma_\phi$ from in Eq.\,\ref{eqn:sigmapost}
are not as easily avoided, one can at least determine the posterior variance of 
$f(\xi)=\phi\ui 0$ by solving 
$\tilde P_{\phi}\bm c=(1,0,\dots,0)'$ for $\bm c$ and using 
$(\tilde P_\phi^{-1})\ui{0,0}=c\ui 0$ in the integral in Eq.\,\ref{eqn:sigmapost}
to determine $\hat\Sigma_\phi\ui{0,0}$.
If the argument errors $\bm\gamma$ are not known to be zero, the integrals over $\bm\gamma$
will usually have to be evaluated numerically even when $\bm\gamma$ is Gaussian as well,
since the integrand is a nonlinear function of $\bm\gamma$.
If $\bm\gamma$ has small variance, 
it can however be feasible to approximate the integrals 
by using a quadratic or 4th-order approximation of $\bm{\tilde\mu}_{\phi}(\bm\gamma)$,
which will be explored in a separate article.

\subsection{Equivariance properties}

In addition to the obvious translation invariance in all dimensions,
the \motabar\ estimator is equivariant under various linear scaling transformations: 
(i) When $\Sigma_\eps$, $\Sigma_\phi$, and $\Sigma_\psi$ are all multiplied with the same constant $C>0$,
then $\bm{\hat\phi}$ remains unchanged and $\hat\Sigma_\phi$ is multiplied by $C$ as well.
(ii) When for all $i$, all $\alpha$, and some $k$, 
$\xi_k$, $x\ui i_k$, and $(\mu_\gamma\ui i)_k$
are all multiplied with the same constant $C>0$,
the distribution of $\bm\gamma$ is stretched by a factor of $C$ along the $k$-th axis, 
and when $\mu_\phi\ui\alpha$, $\mu_\psi\ui{\alpha i}$,
the $\alpha$-th row and column of $\Sigma_\phi$,
and the $\alpha i$-th row and column of $\Sigma_\psi$ are all divided by $C^{\alpha_k}$,
then also $\bm{\hat\phi}\ui\alpha$
and the $\alpha$-th row and column of $\hat\Sigma_\phi$ are divided by $C^{\alpha_k}$
for all $\alpha$.

\section{Special and limit cases}

\begin{figure}\begin{centering}%
\includegraphics[width=0.49\textwidth]{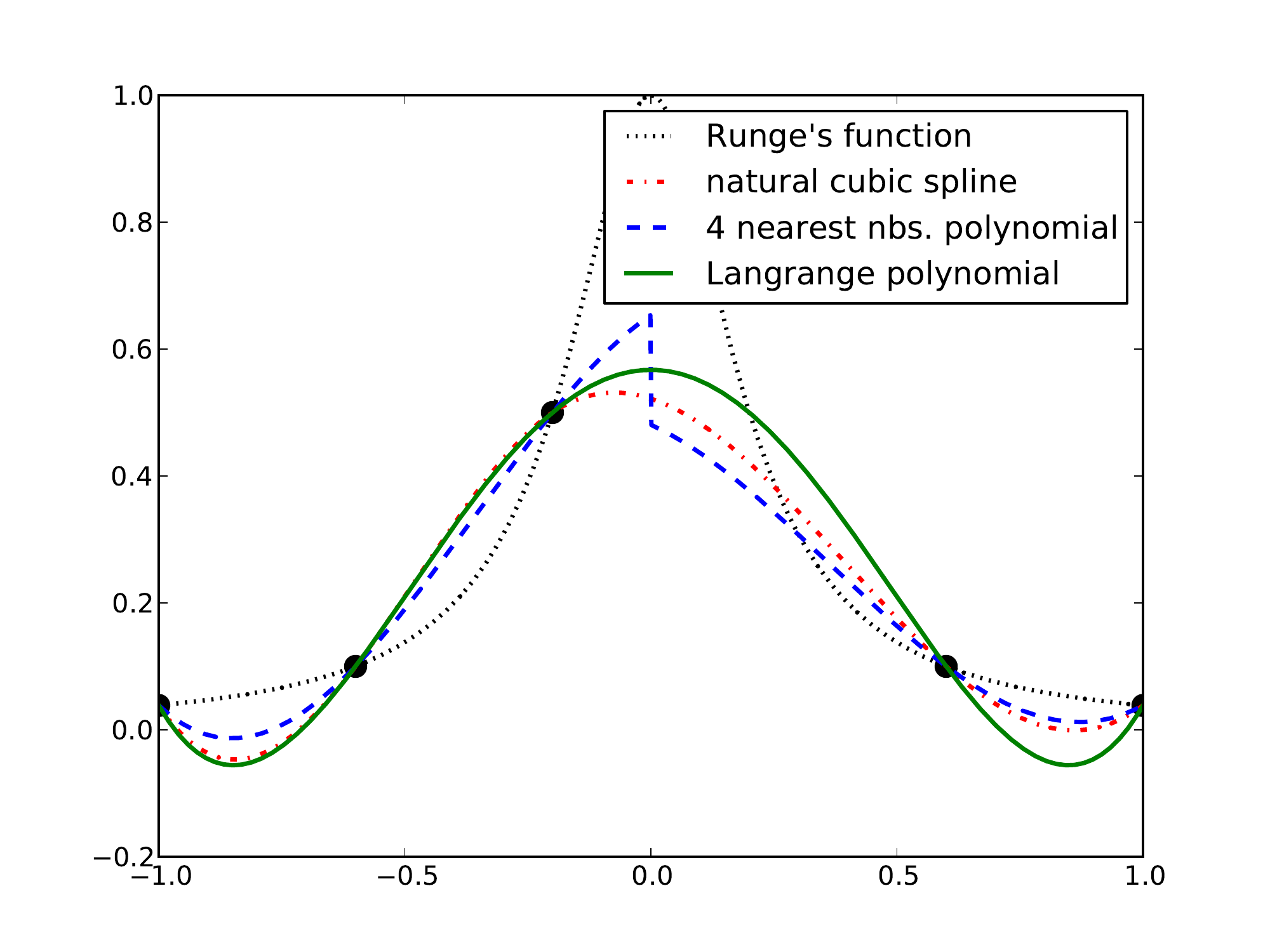}%
\includegraphics[width=0.49\textwidth]{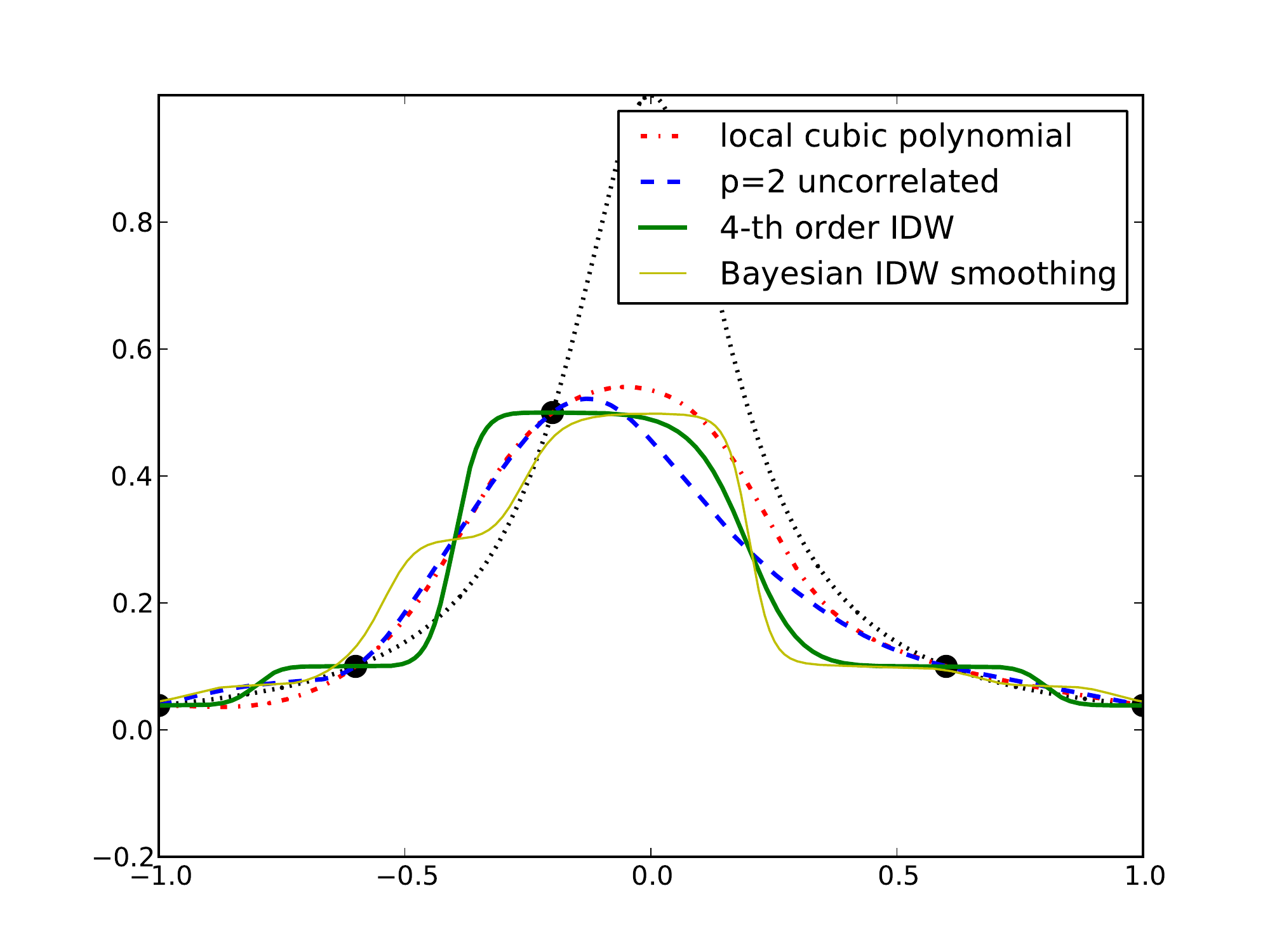}%
\end{centering}
\caption{\label{fig:interpol12}
(Colour online) 
Illustration of special and limit cases of \motabar\ interpolation
for an equidistant sample (black dots)  
from Runge's function $1/(1+25x^2)$ with one missing measurement at 0.2.
Left: Spline interpolation compared to 
\motabar\ cases \ref{sec:nnbs} (piecewise polynomial based on four nearest neighbours) 
and \ref{sec:lagrange} (Lagrange polynomial).
Right: \motabar\ cases \ref{sec:localpoly} (local cubic polynomial regression interpolation), 
\ref{sec:p2uncorr} ($p=2$ with uncorrelated errors and priors), 
\ref{sec:idw4} (4-th order IDW), 
and a smoothing case ($p=5$, small error, penalised intermediate derivatives),
all resulting in nonpolynomial rational functions. 
}\end{figure}

To better understand the effects of the various inputs to \motabar,
let us consider a number of special and limit cases, 
some of which turn out to be equivalent to well-known existing methods,
whereas others are presented to show limitations of our approach. 
Fig.\,\ref{fig:interpol12} illustrates the diversity of results one can get from the same data.

\subsection{Vanishing argument errors} 

For vanishing argument errors, $\rho(\bm\gamma|\bm x)$ becomes a Dirac delta function, 
so that
\begin{equation}\label{eqn:gamma0}
    \bm{\hat\phi} = \bm{\tilde\mu}_\phi(\bm\gamma= 0),\quad
    \hat\Sigma_\phi = \tilde P_\phi(\bm\gamma= 0)^{-1}/2,
\end{equation}
hence the estimate is a pole-free rational function of $\xi$
of degree at most $(2p^2N,2p^2N)$.
For most of the remainder of this section, 
we only derive $\tilde P_{\phi}=\tilde P_{\phi}(\bm\gamma)$ and
$\bm{\tilde\mu}_{\phi}=\bm{\tilde\mu}_{\phi}(\bm\gamma)$ which can then be plugged
into Eqs.\,\ref{eqn:estimator} and \ref{eqn:sigmapost} if argument errors exist.

\subsection{Noninformative priors} 

If there is no prior information about $f$,
an improper $\phi$-prior with $P_\phi\to 0$ 
and a centralised $\psi$-prior with $\bm\mu_\psi\to 0$ can be used,
in which case
\begin{equation}\label{eqn:improper}
    \tilde P_{\phi} \to X' W X,\quad
    \bm{\tilde\mu}_{\phi} \to (X' W X)^{-1} X'W\bm y.
\end{equation}
In this case, $\tilde{\bm\mu}_\phi$ is that value of $\bm\phi$ 
which minimises the quadratic function
\begin{equation}
    L(\bm\phi) = ||A\bm\phi-\bm b||_2^2
    \quad\text{with}\quad
    A = W^{1/2}X,\quad
    \bm b = W^{1/2}\bm y,
\end{equation}
which provides an alternative way of numerical computation.
Although this is formally a weighted least-squares estimator,
it does not estimate a global set of parameters $\bm\beta$ 
as in a global linear regression model $y = X\bm\beta + \bm\eps$, 
but is a local model in which $X$ and $W$ depend on $\xi$,
so that the terms in the estimator have to be computed for each $\xi$ separately.

\paragraph{Consistent estimation of polynomial components.}
If $P_\phi = 0$ and $\bm\mu_\psi = 0$, we have $\tilde P_\phi^{-1}X'WX=I$.
Hence if $\xi=0$ and $\bm y$ is a monomial with $y\ui{i}=(x\ui{i})^\alpha/\alpha!$ for some $\alpha$ with $|\alpha|<p$,
then $\bm y$ equals the $\alpha$-th column of $X$ and hence
$\tilde\mu_\phi = \tilde P_\phi^{-1}X'W\bm y$ equals the $\alpha$-th column of $I$. 
This means that the estimate 
$\tilde\mu_\phi\ui{\beta}$ is zero for $\alpha\neq\beta$ and one for $\alpha=\beta$,
which are the correct derivatives of the given monomial at $\xi=0$.
Consequently, \motabar\ with improper priors is also equivariant under the addition of polynomials,
\eg, a quadratic trend in a time series,
a property shared with numerical differentiation via discrete Legendre polynomials.
However, while the latter has $\hat\phi=C(\bm x,\xi)\bm y$ with an orthogonal coefficient 
matrix $C(\bm x,\xi)$, the \motabar\ coefficient matrix $\tilde P_\phi^{-1}X'W$ is not orthogonal in general.

For proper priors with $P_\phi \neq 0$ and $\bm\mu_\phi=0$, the absolute value of the $\alpha$-th derivative is underestimated
since the prior drags the estimate towards its mean at zero,
and the absolute value of the other derivatives is slightly overestimated (nonzero).
This effect becomes smaller with increasing $p$. 

\subsection{Vanishing value errors: interpolation}

For vanishing value errors, we have $\Sigma_\eps\to 0$,
hence $W\to P_r=\Sigma_r^{-1}$ and
\begin{equation}\label{eqn:eps0}
    \tilde P_\phi \to P_\phi + X' P_r X,\quad
    \bm{\tilde\mu}_\phi \to \tilde P_{\phi}^{-1}\left\{P_\phi\bm\mu_\phi + X' P_r (\bm y-\bm\mu_r)\right\}.
\end{equation} 
In the limit case of $\Sigma_\eps=0$, this can also be derived directly from Eq.\,\ref{eqn:bayes} 
by substituting $\bm r=\bm y-X\bm\phi$.
In that case, however, $\Sigma_r$ can become singular 
when $\xi\to\chi\ui{i}$ for some $i$, which has to be taken care of in the
numerical solution, \eg, by assuming very small but nonzero $\Sigma_\eps$,
or by using the following exact solution for the case $\xi = \chi\ui{i}$:
In that case, $\phi\ui 0=y\ui i$ and,
using the notation $M_{(j,k)},\bm v_{(j)}$ for a matrix $M$ without its $j$th row and $k$th column
and a vector $\bm v$ without its $j$th element,
\begin{align}
    \tilde P_{\phi,(0,0)} &= Q + X_{(i,0)}'\Sigma_{r,(i,i)}^{-1}X_{(i,0)},\\
    \bm{\tilde\mu}_{\phi,(0)} &=
       \tilde P_{\phi,(0,0)}^{-1}\left\{Q\bm\nu + X_{(i,0)}'(\Sigma_{r,(i,i)})^{-1}
       (\bm y_{(i)} - y\ui i \bm e - \bm\mu_{r,(i)})\right\},
\end{align}
where $Q,\bm\nu$ are the prior precision and mean of $\bm\phi_{(0)}$ conditional on $\phi\ui 0=y\ui i$,
and $\bm e=(1,\dots,1)'$ is a vector of ones.

\subsection{Vanishing value errors with noninformative priors:\\ local polynomial regression interpolation}
\label{sec:localpoly}

If in addition to $\Sigma_\eps = 0$ we have $P_\phi = 0$ and $\bm\mu_\psi = 0$,
we get
\begin{equation}\label{eqn:eps0improper}
    \tilde P_\phi = X' P_r X,\quad
    \bm{\tilde\mu}_\phi = (X' P_r X)^{-1}X' P_r \bm y
\end{equation} 
for general $\xi$, while for $\xi=\chi\ui i$, we get 
$\phi\ui 0=y\ui i$, $\tilde P_{\phi,(0,0)} = X_{(i,0)}'VX_{(i,0)}$, and
\begin{align}
    \bm{\tilde\mu}_{\phi,(0)} &=
       (X_{(i,0)}'VX_{(i,0)})^{-1}X_{(i,0)}'V(\bm y_{(i)} - y\ui i\bm e)
\end{align}
with $V=\Sigma_{r,(i,i)}^{-1}$. 
This is interpolation based on local polynomial regression in which the weights
$P_r\ui{i,j}$ decrease as a power of a quadratic form of $\chi\ui{i}-\xi$ and $\chi\ui{j}-\xi$
(\eg, the dash-dotted red line in Fig.\,\ref{fig:interpol12} right). 
Hence, in contrast to other local polynomial methods such as LOESS, 
far away observations get a positive though small weight.

\subsection{Diverging errors}

At the other end of the error scale, for $P_\eps\to 0$, we get $W\to 0$ and thus 
$\tilde P_\phi\to P_\phi$ and $\bm{\tilde\mu}_\phi\to \bm\mu_\phi$.
So if the $\phi$-prior is proper, the posterior approximates it, otherwise it diverges.
The same behaviour obtains for a decreasingly informative $\psi$-prior, \ie, for $P_\psi\to 0$.
This shows that while a noninformative $\phi$-prior might be chosen,
the $\psi$-prior must be proper since it regulates the overall variability of the estimate.

\subsection{Penalised higher derivatives} 

If the value error variance is finite ($\Sigma_\eps\neq 0$), 
one might, on the other hand, assume that the derivatives of $f$ of degree $\geq p$ are negligible
compared to the scale of $\bm\eps$, 
\eg, because $f$ is believed to be approximately a polynomial of degree at most $p-1$.
Then, taking the limit $\bm\mu_\psi\to 0$ and $\Sigma_\psi\to 0$,
we get 
\begin{equation}\label{eqn:psi0}
    \tilde P_{\phi} \to P_\phi + X' P_{\eps} X,\quad
    \bm{\tilde\mu}_{\phi} \to \tilde P_{\phi}^{-1}\left(P_\phi\bm\mu_\phi + X' P_{\eps} \bm y\right),
\end{equation}
which is a form of Bayesian polynomial regression.
Although the squared weight matrix $W=P_\eps$ no longer depends on $\xi$ here, 
this is still not a global regression method since the $\phi$-prior might dependent on $\xi$,
\eg, because one assumes some linear or nonlinear trend or some change in variability with $\xi$.

\subsection{Penalised higher derivatives with a noninformative prior:\\ ordinary global polynomial regression}

A truly global regression method can be obtained 
by using the noninformative $\phi$-prior with $P_\phi\to 0$
in addition to $\bm\mu_\psi\to 0$ and $\Sigma_\psi\to 0$.
If $N\ge{p-1+d\choose d}$, then
\begin{equation}\label{eqn:psi0improper}
    \tilde P_\phi \to X' P_\eps X,\quad
    \bm{\tilde\mu}_\phi \to (X' P_\eps X)^{-1}X' P_\eps \bm y.
\end{equation} 
In the limit, this is just ordinary global least-squares polynomial regression
with polynomials of order $p-1$ and possibly correlated value errors of differing magnitude. 

\subsubsection{Linear regression with argument errors vs.~total least squares}
For $d=1$, $p=2$, $\phi=(a,b)'$, and nonvanishing argument measurement errors,
our model becomes the ``errors in the variables'' linear regression model
\begin{equation}
    y\ui i = a + b(x\ui i-\gamma\ui i) + \eps\ui i.
\end{equation}
Since this is linear in $\gamma_i$, the integral over $\bm\gamma$ in Eq.\,\ref{eqn:bayes}
can be solved analytically.
For $\xi=0$, $P_\phi=0$, $\bm\mu_\psi=0$, $\Sigma_\psi=0$, $\Sigma_\eps=\sigma^2_\eps I$, 
and $\Sigma_\gamma=\sigma^2_\gamma I$, this results in
the non-Gaussian posterior
\begin{align}
    \rho(a,b|\bm x,\bm y) 
    &\propto \int d^N\bm\gamma\,\exp\Big(
        -||\bm\gamma||_2^2/2\sigma_\gamma^2
        -||\bm y-a\bm e-b\bm x+b\bm\gamma||_2^2/2\sigma_\eps^2\Big)\nonumber\\
    &\propto (\sigma_\eps^2+b^2\sigma_\gamma^2)^{-N/2} 
        \exp\Big\{-||\bm y-a\bm e-b\bm x||_2^2/2(\sigma^2_\eps+b^2\sigma_\gamma^2)\Big\}.
\end{align}
The posterior mode has $a = \bar y - b \bar x$ and
\begin{equation}
    N\sigma_\gamma^4 b^3 + \sigma_\gamma^2 s_{xy}b^2 
    + (N\sigma_\eps^2\sigma_\gamma^2 + \sigma_\eps^2 s_{xx} - \sigma_\gamma^2 s_{yy})b - \sigma_\eps^2 s_{xy} = 0,
\end{equation} 
where 
$\bar x=\sum_i x\ui i/N$ and
$\bar y=\sum_i y\ui i/N$,
$s_{xx}=\sum_i(x\ui i-\bar x)^2$,
$s_{yy}=\sum_i(y\ui i-\bar y)^2$, and
$s_{xy}=\sum_i(x\ui i-\bar x)(y\ui i-\bar y)$.
If $s_{xy}>0$, $b$ is the largest solution of the above equation,
otherwise the smallest, and it has the same sign as $s_{xy}$.

When the same model is estimated with the {\em total least-squares} method (aka {\em Deming regression})
studied already by \citet{Kummell1879}, 
the equation is
\begin{equation*}
    \sigma_\gamma^2 s_{xy}b^2 
    + (\sigma_\eps^2 s_{xx} - \sigma_\gamma^2 s_{yy})b - \sigma_\eps^2 s_{xy} = 0,
\end{equation*}
instead, which is symmetric under an exchange of ``dependent'' and ``independent'' variables 
and leads to a larger absolute value of $b$. 
For $\sigma_\gamma^2\ll s_{xx}/N$, this difference vanishes,
and for $\sigma_\gamma\to 0$, both solutions converge to 
the ordinary least-squares linear regression line with
$b\to s_{xy}/s_{xx}$.
For $\sigma_\eps\to 0$, total least squares gives $b\to s_{yy}/s_{xy}$ independently of $\sigma_\gamma$,
while the cubic equation gives 
\begin{align*}
b\to \left\{\sgn(s_{xy})\sqrt{4N\sigma_\gamma^2 s_{yy}+s_{xy}^2}-s_{xy}\right\}/2N\sigma_\gamma^2,
\end{align*}
which still depends on $\sigma_\gamma$
and has $b\approx\sgn(s_{xy})\sqrt{s_{yy}/N}/\sigma_\gamma\to 0$ for $\sigma_\gamma\to\infty$.
This comparison shows that unlike in total least squares, 
it is essential in \motabar\ whether we consider $y$ a function of $x$ or vice versa.

To understand why the effects of value and argument errors are different in the linear model 
and why the cubic equation is not symmetric in $x,y$, 
consider the simple case where the measurements are $\bm x=\bm y=(-2,0,2)$ 
and only the middle one, $(x\ui 2,y\ui 2)=(0,0)$, has errors in both dimensions
which are independent with $\gamma\ui 2,\eps\ui 2\in\{1,-1\}$ with equal probability.
Then the real data $\bm\chi,f(\bm\chi)$ are either 
$(-2,-1,2),(-2,-1,2)$ or $(-2,1,2),(-2,1,2)$, both giving slope $b=1$ in linear regression,
or $(-2,-1,2),(-2,1,2)$ or $(-2,1,2),(-2,-1,2)$, both giving slope $b\approx 0.84$,
so that the \motabar\ posterior mean is $b\approx 0.92$,
whereas total least squares results in $b=1$ because of the obvious symmetry.

\subsection{Larger order with non-informative prior: Lagrange interpolation}
\label{sec:lagrange}
For $d=1$, $N=m=p$, and $P_\phi=0$, 
we get Lagrange interpolation with $\bm{\tilde\mu}_\phi = X^{-1}\bm y$
(\eg, the solid green line in Fig.\,\ref{fig:interpol12} left). 
For $d>1$ and $N=m$, it depends on the placement of the $\chi\ui i$ 
whether the resulting polynomial is an interpolant or not (see \citet{Gasca2001} for an overview).

\subsection{Minimal order with uncorrelated errors and priors: Bayesian IDW smoothing of order two}

A particularly simple case occurs for the minimal choice of $p$, $p=1$,
and when both value errors and priors are uncorrelated, 
so that $\Sigma_\eps$ and $\Sigma_\psi$ are diagonal, 
and $\Sigma_\phi=(\sigma^2_\phi)$.
Then we have $\bm\phi = (f(\xi))$, $X = (1,\dots,1)'$, and
$W\ui{i,j} = \delta_{ij}w\ui{i}$, where
\begin{equation}\label{eqn:bidww}
    w\ui{i} = \frac{1}{s\ui{i}+\Sigma_\eps\ui{i,i}},\quad
    s\ui{i} = \sum_{k=1}^d\Sigma_\psi\ui{ki,ki}(\chi\ui{i}_k-\xi_k)^2.
\end{equation} 
The latter is the squared distance between $\chi\ui i$ and $\xi$
as measured by the quadratic norm that is weighted with the prior derivative variances $\Sigma_\psi\ui{k,k}$.
The posterior distribution of $f(\xi)$ given $\bm\gamma$ is then Gaussian with
mean and variance given by   
\begin{equation}\label{eqn:bidw}
    \tilde\mu_{\phi} =  
       \frac{\mu_\phi + 
           \sigma^{2}_\phi\sum_i w\ui{i}(y\ui{i} - \mu_r\ui{i})}%
           {1 + \sigma^{2}_\phi\sum_i w\ui{i}},\quad
    \tilde\sigma_{\phi}^2 = \frac{\sigma^{2}_\phi}{1 + \sigma^{2}_\phi\sum_i w\ui{i}}.
\end{equation}
This is a generalization of inverse distance weighting
that takes into account value error via the occurrence of $\sigma_\eps\ui i$ in Eq.\,\ref{eqn:bidww}
and prior information via the occurrence of $\mu_\phi,\sigma_\phi^2$ in Eq.\,\ref{eqn:bidw}.
We propose to call this method {\em Bayesian IDW smoothing.}
Note that, as expected, $\tilde\mu_{\phi}$ is a rational function of $\xi$ of degree 
at most $(2p^2N,2p^2N)=(2N,2N)$.

\subsection{Minimal order with vanishing errors and noninformative priors:\\ ordinary IDW with squared distances}

Setting $\Sigma_\eps=0$, $P_\phi=0$, and $\Sigma_\psi = \sigma_\psi^2 I$ in the preceding,
we get ordinary IDW with squared distances,
\begin{equation}\label{eqn:idw}
    w\ui{i} = \frac{1}{\sigma_\psi^2||\chi\ui{i}_k-\xi_k||_2^2},\quad
    \tilde\mu_{\phi} = \frac{\sum_i w\ui{i}y\ui{i}}{\sum_i w\ui{i}},\quad
    \tilde\sigma_{\phi}^2 = \frac{1}{\sum_i w\ui{i}},
\end{equation} 
where $\tilde\mu_{\phi}$ is now a rational function of $\xi$ of degree exactly $(2N-2,2N-2)$.
Note that while $\tilde\mu_{\phi}$ is independent from $\sigma_\psi^2$,
the posterior variance is proportional to $\sigma_\psi^2$.
In other words, our Bayesian approach shows that the uncertainty of the IDW estimate of order two
is directly proportional to the scale of the derivatives of $f$.

\subsection{Order two in one dimension with uncorrelated errors and priors}
\label{sec:p2uncorr}

If $p=2$, $d=1$, $\bm\mu_\phi=0$, $\bm\mu_\psi=0$, $\Sigma_\eps$ and $\Sigma_\psi$ are diagonal, 
and $P_\phi=\diag(a,b)$, 
then
\begin{align}
    \bm{\tilde\mu}_\phi &= \frac{\left(\begin{array}{l}
                \sum_i z\ui i y\ui i\\
                \sum_i w\ui i \{a(\chi\ui i-\xi) - \sum_j w\ui j(\chi\ui j-\chi\ui i)\} y\ui i
            \end{array}\right)
    	}{a\{b+\sum_i w\ui i(\chi\ui i-\xi)^2\} + \sum_i z\ui i},\\
    \tilde P_\phi &= \left(\begin{array}{ll}
                a + \sum_i w\ui{i}& \sum_i w\ui{i}(\chi\ui{i}-\xi)\\
                \sum_i w\ui{i}(\chi\ui{i}-\xi) & b + \sum_i w\ui{i}(\chi\ui{i}-\xi)^2
            \end{array}\right),\\
    z\ui i &= w\ui i\{b + \textstyle\sum_j w\ui j(\chi\ui j-\xi)(\chi\ui j-\chi\ui i)\},\\
    w\ui i &= \frac{1}{\Sigma_\psi\ui{i,i}(\chi\ui{i}-\xi)^4/2 + \Sigma_\eps\ui{i,i}}.
\end{align}
Note that $\tilde\mu_{\phi}\ui 0$ 
does not only depend on the distances of the $\chi\ui i$ from $\xi$ as encoded in $w\ui i$, 
but also on the relative position of $\chi\ui i$ and $\chi\ui j$, via the mixture terms in $z\ui i$.
See the dashed blue line in Fig.\,\ref{fig:interpol12} right for an example. 
This dependency vanishes if we let $a\to 0$ and $b\to\infty$, where we get an instance of the following case:

\subsection{Uncorrelated errors, penalised intermediate derivatives, and specific priors:\\ IDW smoothing}
\label{sec:idw4}

Assume that $\bm\mu_\phi=0$, $\bm\mu_\psi=0$, $\Sigma_\eps$ is diagonal,
$P_\phi=\diag(a,b,\dots,b)$,
and $\Sigma_\psi\ui{\alpha i,\beta j}=\delta_{\alpha\beta}\delta_{ij}v\ui i$.
If we let the prior for $\phi\ui 0$ become noninformative by letting $a\to 0$, 
but let the prior for $\phi\ui\alpha$ with $|\alpha|>0$ become sharp at $\phi\ui\alpha=0$ by letting $b\to\infty$,
then
\begin{equation}
    \tilde\mu_\phi^0 \to \frac{\sum_i w\ui i y\ui i}{\sum_i w\ui i}\quad\text{with}\quad
    w\ui i = \frac{1}{v\ui i||\chi\ui{i}-\xi||_2^{2p}/p! + \Sigma_\eps\ui{i,i}}.
\end{equation}
In other words, the \motabar\ estimate converges to the ordinary IDW smoothing (or interpolation, if $\Sigma_\eps=0$)
solution with exponent $2p$.
\Eg, $p=2$ and $\Sigma_\eps=0$ gives 4-th IDW interpolation
(solid green line in Fig.\,\ref{fig:interpol12} right).
Note that this, however, also shows that using IDW with higher powers than two corresponds to 
implicitly assuming that some derivatives vanish which the result shows not to vanish after all.
Hence the only plausible form of IDW is the one with squared distances.      

\paragraph{Inconsistent derivatives.}
This example also illustrates an unintuitive feature of \motabar:
The estimates derivative $\tilde\mu_\phi\ui\alpha$ need not coincide 
with the derivative of the estimated value $D^\alpha\tilde\mu_\phi\ui 0$ with respect to $\xi$.
In IDW smoothing, the estimated slope is zero as enforced by the sharp prior,
although the estimated value is not constant.

\subsection{Large order with penalised intermediate derivatives: piecewise estimators}
\label{sec:nnbs}

On the upper end of the order spectrum, 
for certain choices of priors,
one can choose very large orders $p$ without running into numerical infeasibilities.
A special limit case obtains with similar priors as above:
Assume $\Sigma_\eps=0$, 
a noninformative prior for the function value $f(\xi)=\phi\ui{0}$ (so that $P_\phi\ui{0,0}=0$),
an uncorrelated and centralised $\psi$-prior with $\bm\mu_\psi=0$ and $\Sigma_\psi\ui{\alpha i,\beta j}=\delta_{\alpha\beta}\delta_{ij}v\ui i$,
and that all derivatives of $f$ of order $1\dots p-1$ vanish at $\xi$ 
(so that $\mu_\phi\ui{\alpha}=0$ and $\Sigma_\phi\ui{\alpha,\alpha}=0$ for $|\alpha|>0$).
In the limit for $p\to\infty$,
then only the nearest neighbour $\chi\ui{i}$ of $\xi$ is used in the estimation.
This is because then $P_r$ is diagonal and $P_r\ui{j,j}/P_r\ui{i,i}\to 0$ for $j\neq i$, so that
$\rho(\phi\ui{0}|\bm x,\bm y,\bm\gamma)\sim\exp\{-P_r\ui{i,i}(\phi\ui{0}-y\ui{i})^2\}$.
In other words, \motabar\ then estimates $f(\xi)$ by $\hat\phi\ui{0}\to y\ui{i}$,
as in simple {\em nearest neighbour estimation}.
That is, for large $p$ and these priors, 
\motabar\ estimate approximates a step function 
that is constant on the Voronoi cells (aka Thiessen polyhedrons) of the sample.
As the 4-th order IDW example in Fig.\,\ref{fig:interpol12} (right) shows,
the step-like shape can already be seen for small values of $p$.
A step-like function with a similar smoothening at the edges also obtains for $p\to\infty$ 
when argument errors $\bm\gamma$ are not vanishing, since then 
$\hat f(\xi)\to\sum_i \prob(\forall j:||\chi\ui i-\xi||\leq||\chi\ui j-\xi||)y\ui i$.
On the other hand, nonvanishing value errors, even if very small, 
can lead to additional levels at the means of more than one nearest neighbour,
showing that this limit behaviour is quite unstable
(see, \eg, the thin yellow line in Fig.\,\ref{fig:interpol12} right).    

One can also get higher-order piecewise polynomial estimates,
by assuming that only the derivatives of $f$ of order $q+1\dots p-1$ vanish at $\xi$
for some $q<p$,
and using an improper prior with $P_\phi\ui{\alpha,\alpha}=0$ for $|\alpha|\leq q$.
Then the \motabar\ estimate approximates a piecewise polynomial interpolant of degree $q$ 
that uses the $d+q\choose d$ observations nearest to $\xi$ (assuming those are in general position).
For $q=1$, this leads to a piecewise linear estimate that need not, however, 
coincide with ordinary linear interpolation since, \eg, 
the two nearest neighbours of $\xi\in\reals$ might both lie left of $\xi$,
leading to a discontinuity in the estimate to the right of $\xi$. 
Only when the positions $\chi\ui i$ build a regular polyhedral grid, 
the estimate approximates ordinary linear interpolation.
For $d=1$, this requires equidistant measurement positions,
and for $d=2$ a regular triangular grid is required,
whereas on a square grid, one gets a discontinuous linear interpolation
with four square interpolation cells per grid cell, 
each corresponding to a different choice of three of the four corners of the grid cell. 
Analogous results hold for higher-dimensional rectangular grids.
For $q=3$ and $d=1$, the limit result is a piecewise cubic polynomial which is,
however, not a cubic spline since although its 2nd derivative is continuous, 
its value and slope are not
(\eg, the dashed blue line in Fig.\,\ref{fig:interpol12} left). 


\subsection{Extreme arguments of interest}

If the distance between $\xi$ and the data positions $\chi\ui i$ diverges,
\ie, if $||\xi-\chi\ui i||_2\to\infty$, we have $X'WX\to 0$.
If the $\phi$-prior is informative ($P_\phi\neq 0$), we then have
\begin{equation}\label{eqn:extremexi}
    \tilde P_\phi \to P_\phi,\quad
    \bm{\tilde\mu}_\phi \to \bm\mu_\phi,
\end{equation} 
that is, the data are too far away to drag the posterior significantly away from the prior.
With a noninformative $\phi$-prior, on the other hand,
the estimate for $||\xi-\chi\ui i||_2\to\infty$
will either approximate the sample mean (if $p=1$)
or will diverge to $\pm\infty$ (almost surely for $p>1$).

\subsection{Large amounts of data: conjectured exponential rate of convergence}

If the $\chi\ui i$ are either regularly distributed
or drawn from a sufficiently smooth distribution,
the distance between $\xi$ and the closest $\chi\ui i$ 
will decay at a rate $\sim N^{-1/d}$ for $N\to\infty$.
We conjecture that then the posterior precision grows 
at a rate
\begin{align}\label{eqn:conjecture}
    \tilde P_\phi\ui{\alpha,\beta} &\sim N^{1+(2p-|\alpha|-|\beta|-1)/d} &\text{if~} & \Sigma_\eps=0 &\text{(conjectured)}\\
    \tilde P_\phi\ui{\alpha,\beta} &\sim N &\text{if~} & \Sigma_\eps\neq 0 &\text{(conjectured)}
\end{align}
for nonpathological error distributions and priors.
The rationale for this conjecture is this:
Assume that $\Sigma_\eps=0$, $\Sigma_\psi=\sigma_\psi^2 I$, 
and the $N$ arguments $\chi\ui i$ are sufficiently uniformly distributed 
over a $d$-dimensional ball of unit radius about $\xi$,
and let $|\alpha|,|\beta|<p$.
Then
\begin{equation*}
    \tilde P_\phi\ui{\alpha,\beta}-P_\phi\ui{\alpha,\beta}
        \propto\textstyle\sum_{i=1}^N (\chi\ui i-\xi)^{\alpha+\beta}||\chi\ui i-\xi||_2^{-2p},
\end{equation*}
and this sum is asymptotically proportional to the integral
\begin{align*}
    & N\int_{||x-\xi||_2\in[N^{-1/d},1]} d^dx\,(x-\xi)^{\alpha+\beta}||x-\xi||_2^{-2p}\\
    & \sim N\int_{z=N^{-1/d}}^1 dz\,z^{|\alpha|+|\beta|}z^{-2p}
    \sim N^{1+(2p-|\alpha|-|\beta|-1)/d}.
\end{align*}
The conjecture implies $\tilde\Sigma_\phi\ui{\alpha,\beta}\sim 1/N^{1+(2p-|\alpha|-|\beta|-1)/d}$,
in particular $|\hat\phi\ui\alpha-(D^\alpha f)(\xi)|\sim N^{(1/2+|\alpha|-p)/d-1/2}$.
In particular, for $d=1$, the conjectured rate of convergence of $\hat\phi\ui 0$ to $f(\xi)$ 
is $N^p$, the same as for spline interpolation.

\section{Comparison with Wang et al.'s approach}
\label{sec:wang}

\citet{Wang2010c} introduced a different approximation scheme that is also based on a local Taylor expansion. 
Adapting their terminology and notation to ours, that scheme can be described as defining an estimator 
\begin{align}
	\hat f_\wang(\xi) = \bm a'\bm y
\end{align}
for $f(\xi)$, where the weights $a_i$ are chosen to minimise the squared norm
\begin{align}
	||\bm a||^2_\wang &= \bm a'(X V_\phi X' + V_y)\bm a
\end{align} 
subject to a number of constraints 
\begin{align}
	\textstyle\sum_i a\ui{i} &= 1, & \textstyle\sum_i a\ui{i}X\ui{i,\alpha} &= 0
\end{align}
for all $\alpha$ with $0<|\alpha|<q$ for some $q<p$, 
where $V_\phi$ and $V_y$ are diagonal matrices whose entries are certain parameters $\sigma\ui{\alpha}_\phi$ and error variances,
\begin{align}
	V_\phi\ui{\alpha,\alpha} &= (\sigma\ui{\alpha}_\phi)^2,
	& V_y\ui{i,i} &= (\sigma\ui{i}_r)^2 + (\sigma\ui{i}_\eps)^2.
\end{align} 
Although the parameters $(\sigma\ui{\alpha}_\phi)^2$ correspond to the prior variances in our approach,
Wang et al.'s rationale is not Bayesian.
Their motivation is rather that $||\bm a||^2_\wang$ is an estimator of the squared approximation error $\{\hat f_\wang(\xi)-f(\xi)\}^2$,
and the constraints make sure that the estimator is correct if
the true $f$ is a polynomial of degree $\le q$. 
As we see from the usage of $V_\phi$ and $V_y$ instead of $\Sigma_\phi$ and $\Sigma_y=W^{-1}$,
their scheme implicitly assumes uncorrelated ``priors'' and errors, 
but one can easily adapt it to deal with correlations by using
\begin{align}
	||\bm a||^2_\corr &= \bm a'(X \Sigma_\phi X' + \Sigma_y)\bm a
\end{align} 
instead of $||\bm a||^2_\wang$. 
Also, they do not consider the case of argument errors, 
and for these it does not seem obvious how to deal with them consistently in their framework.

To compare their scheme to ours in the case of no argument errors, note that 
the \motabar\ estimate $\tilde{\bm\mu}_\phi$ is a linear function of $\bm y$
only when the prior for $\bm\phi$ is improper and the prior for $\bm\psi$ is centralised,
while otherwise the estimate is only an {\em affine} function of $\bm y$.
Hence let us assume $P_\phi=0$ and $\bm\mu_\psi=0$ in this section.
In that case, the \motabar\ estimator $\tilde{\bm\mu}_\phi$ itself is that
$\bm\phi$ which minimises the quadratic function
\begin{equation}
	L(\bm\phi) = ||A\bm\phi-\bm b||_2^2
	\quad\text{with}\quad
	A = W^{1/2}X,\quad
	\bm b = W^{1/2}\bm y.
\end{equation}
However, while in Wang et al.'s approach a constrained problem obtains, the above is an unconstrained problem. 
Taking the corresponding limit of $P_\phi\to 0$ in Wang et al.'s scheme, their target function becomes
\begin{align}
	||\bm a||^2_\corr &\approx \bm a'X \Sigma_\phi X'\bm a
\end{align} 
so that their coefficients depend on the relative variances (and correlation structure) of $\bm\phi$
but no longer on $\Sigma_y$, while ours depends on the latter but not on the former.

Despite these differences,
both approaches seem to have the same rate of convergence for $N\to\infty$ 
and reduce to ordinary polynomial regression or IDW for certain choices of control parameters.

\section{Non-Gaussian cases and constraints}

Prior knowledge about $\bm\phi$,
such as constraints of the form $(D^\alpha f)(x)\in[a,b]$
for some $\alpha$ and $a,b\in\reals\cup\{\pm\infty\}$,
can easily be incorporated into \motabar\ estimation 
by choosing a suitable prior distribution for $\bm\phi$,
such as a uniform distribution of $\phi\ui\alpha$ on $[a,b]$
or a Gaussian restricted to this interval.
If one then rewrites the chosen prior in the form
\begin{align}
    \rho(\bm\phi) &\propto \rho_0(\bm\phi)\exp\{-(\bm\phi-\bm\mu_\phi)' P_\phi (\bm\phi-\bm\mu_\phi)\},
\end{align}
which can be seen as a generalization of Eq.\,\ref{eqn:phiprior},
it is easy to see that the resulting posterior simply involves the same factor $\rho_0(\bm\phi)$,
leading to a posterior distribution of
\begin{align}\label{eqn:posteriorgeneralized}
    \rho(\bm\phi|\bm x,\bm y) &\propto \rho_0(\bm\phi)\int d\bm\gamma\rho(\bm\gamma|\bm x)  
       \exp\{-(\bm\phi-\bm{\tilde\mu}_{\phi})' \tilde P_{\phi} (\bm\phi-\bm{\tilde\mu}_{\phi})\}
\end{align}
with the same $\bm{\tilde\mu}_{\phi}$ and $\tilde P_{\phi}$ as before.
However, since the posterior mean is then no longer equal to $\bm{\tilde\mu}_{\phi}$,
one either needs to integrate Eq.\,\ref{eqn:posteriorgeneralized} explicitly
or use the posterior mode instead.

In some cases, the posterior mean can be determined analytically.
\Eg, two-sided constraints of the form $(D^\alpha f)(x)\in[a,b]$
with finite bounds $a,b\in\reals$
can most easily be modelled by putting $P_\phi\ui{\alpha,\alpha}=0$ and 
$\rho_0(\bm\phi)=\Theta(\phi\ui\alpha-a\ui\alpha)\Theta(b\ui\alpha-\phi\ui\alpha)$,
where $\Theta$ is the Heaviside step function with $\Theta(\phi)=1$ for $\phi\geq 0$ and $\Theta(\phi)=0$ otherwise.
For one-sided constraints of the form $(D^\alpha f)(x)\geq a\ui\alpha$,
one can use $\rho_0(\bm\phi)=\Theta(\phi\ui\alpha-a\ui\alpha)$.

\subsection{Function value constrained to an interval}

Using the prior factor  $\rho_0(\bm\phi)=\Theta(\phi\ui 0-a)\Theta(b-\phi\ui 0)$,
the marginal posterior density of $\phi\ui 0$ is proportional to 
\begin{align}
    & \Theta(\phi\ui 0-a)\Theta(b-\phi\ui 0)\exp\{-(\phi\ui 0-\mu)^2/2\sigma^2\}
\end{align} 
with $\mu=\tilde\mu_\phi\ui 0$ and $\sigma^2 = (\tilde P_\phi^{-1}/2)\ui{0,0}$.
For a density of this form, the mean is given by
\begin{align}
    \nu &= \mu + \sigma\frac{\exp\{-(\mu-a)^2/2\sigma^2\}-\exp\{-(\mu-b)^2/2\sigma^2\}}%
        {\erf\{(\mu-a)/\sqrt{2}\sigma\}-\erf\{(\mu-b)/\sqrt{2}\sigma\}}\sqrt{\frac{2}{\pi}} \in (a,b). 
\end{align} 
For $\mu$ outside $[a,b]$, this asymptotically equals 
$b-\sigma(-1/z+2/z^3-10/z^5+74/z^7-706/z^9)$ with $z=(b-\mu)/\sigma$ for $\mu\to\infty$ 
(good approximation for $\mu>b+4\sigma$),
or $a+\sigma(-1/z+2/z^3-10/z^5+74/z^7-706/z^9)$ with $z=(\mu-a)/\sigma$ for $\mu\to-\infty$ 
(good approximation for $\mu<a-4\sigma$). 
With this constraint, the \motabar\ estimator becomes
\begin{align}
    \hat\phi &= \int d\bm\gamma\rho(\bm\gamma|\bm x)\nu,
\end{align}
where $\nu$ depends on $\bm\gamma$ via $\mu$ and $\sigma$.

\subsection{One-sided constraints}

Letting $b\to\infty$ in the preceding, we get
\begin{align}
    \nu &= \mu + \sigma\frac{\exp\{-(\mu-a)^2/2\sigma^2\}}%
        {\erf\{(\mu-a)/\sqrt{2}\sigma\}+1}\sqrt{\frac{2}{\pi}} > a. 
\end{align} 

\subsection{Non-Gaussian errors and quantile regression}

To see that our methodology can be fruitful also in the case of non-Gaussian $\psi$-priors and value errors,
consider the case of $d=1$, vanishing argument errors, an improper $\phi$-prior, 
independent Laplace $\psi$-priors with $\rho(\psi\ui i)\propto\exp(-b\ui i|\psi\ui i|)$ for some $b\ui i>0$,
and independent value errors with asymmetric Laplace distributions
$\rho(\eps\ui i)\propto \exp\{-a\ui i\rho_{\tau\ui i}(\eps\ui i)\}$,
where $\rho_\tau(z)=z\{\tau-\Theta(-z)\}$, $a\ui i>0$, and $\tau\ui i\in [0,1]$.
Then 
\begin{align}\nonumber
	\rho(\bm y|\bm x,\bm\phi) &\propto\prod_i\int dr\ui i\exp
		\{-a\ui i\rho_{\tau\ui i}(y\ui i-(X\bm\phi)\ui i-r\ui i)-b\ui i|\textstyle\frac{r\ui i}{X\ui{i,p}}|\}\\
		&\propto\prod_i g\{y\ui i-(X\bm\phi)\ui i,a\ui i,\tau\ui i,b\ui i/|X\ui{i,p}|\},
\end{align}
where
\begin{align}
	g(z,a,\tau,c) = \left\{\begin{array}{ll}
		\frac{\exp\{za(1-\tau)\}}{c+a(1-\tau)}
		+\frac{\exp\{za(1-\tau)\}-\exp(zc)}{c-a(1-\tau)}
		+\frac{\exp(zc)}{c+a\tau}
		&\text{if~}z\le 0\\
		\frac{\exp(-za\tau)}{c+a\tau}
		+\frac{\exp(-za\tau)-\exp(-zc)}{c-a\tau}
		+\frac{\exp(-zc)}{c+a(1-\tau)}
		&\text{if~}z\ge 0
	\end{array}\right.,
\end{align}
and the posterior mode of $\bm\phi$ given $\bm x$, $\bm y$ is the one that maximises $\rho(\bm y|\bm x,\bm\phi)$.
For $a\ui i\equiv a$, $\tau\ui i\equiv\tau$, and sharp $\psi$-priors with $b\ui i\to\infty$,
this reduces to minimizing the quantile regression loss function,
$\sum_i \rho_\tau\{y\ui i - (X\bm\phi)\ui i\}$,
so the above can be considered a novel form of local polynomial quantile regression.
Similar analytical forms of $g$ obtain also for other forms of $\psi$-priors, \eg, Gaussian or uniform ones.

\section{Choice of control parameters}

\subsection{Consistent derivatives}

With a noninformative $\phi$-prior, the inconsistency between $\tilde\mu_\phi\ui\alpha$ 
and the $\alpha$-th derivative of $\tilde\mu_\phi\ui 0$ with respect to $\xi$
will decrease with $p$ and vanish for $p\to\infty$.
Therefore, if one is interested in all derivatives of order $\leq q$,
we suggest to use a noninformative prior for all $\phi\ui\alpha$ with $|\alpha|\le q$,
and use a $p$ somewhat larger than $q$,
using either informative or noninformative priors for the derivatives of order $>q$.

\begin{figure}\begin{centering}%
\includegraphics[width=0.49\textwidth]{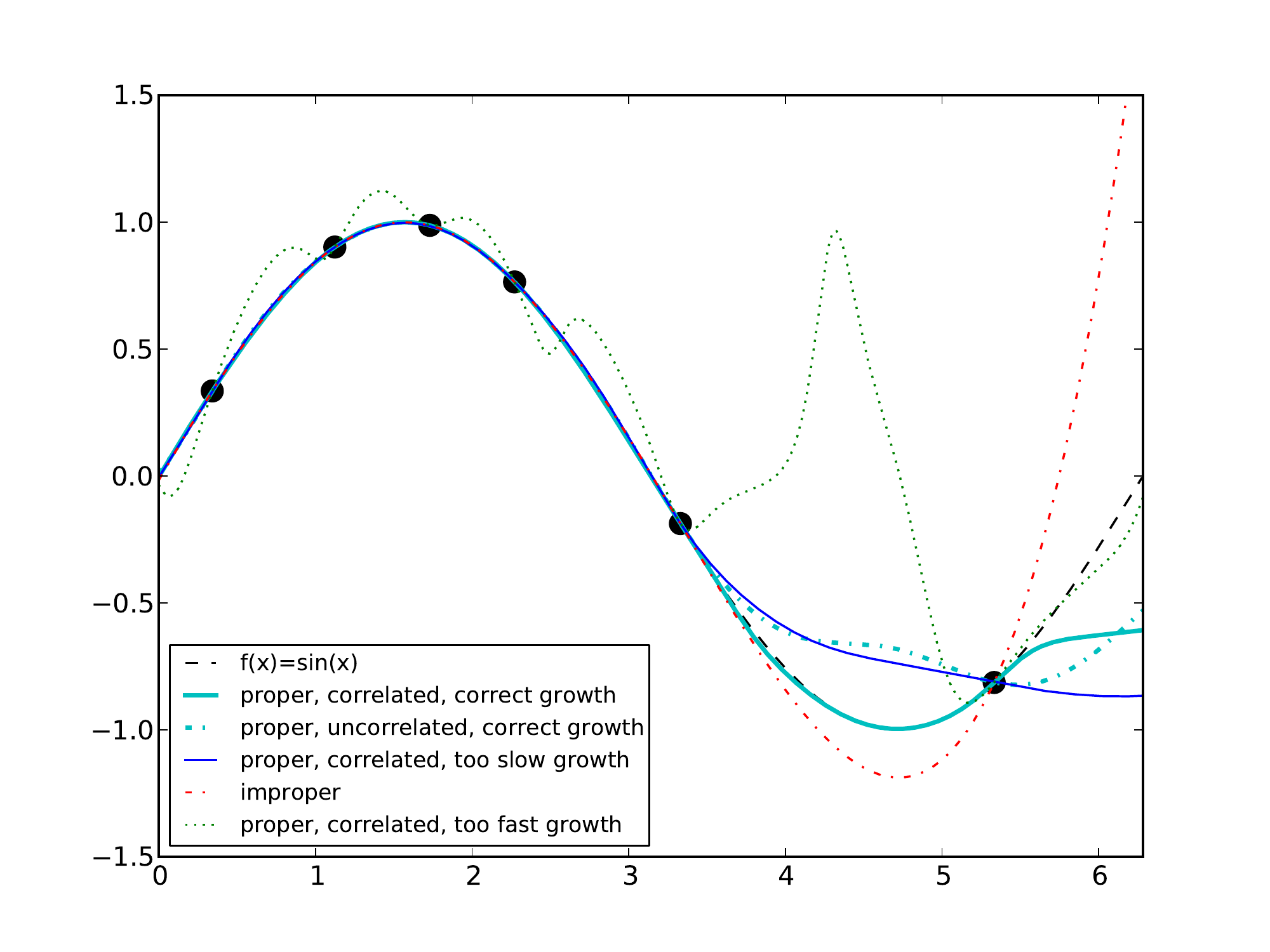}%
\includegraphics[width=0.49\textwidth]{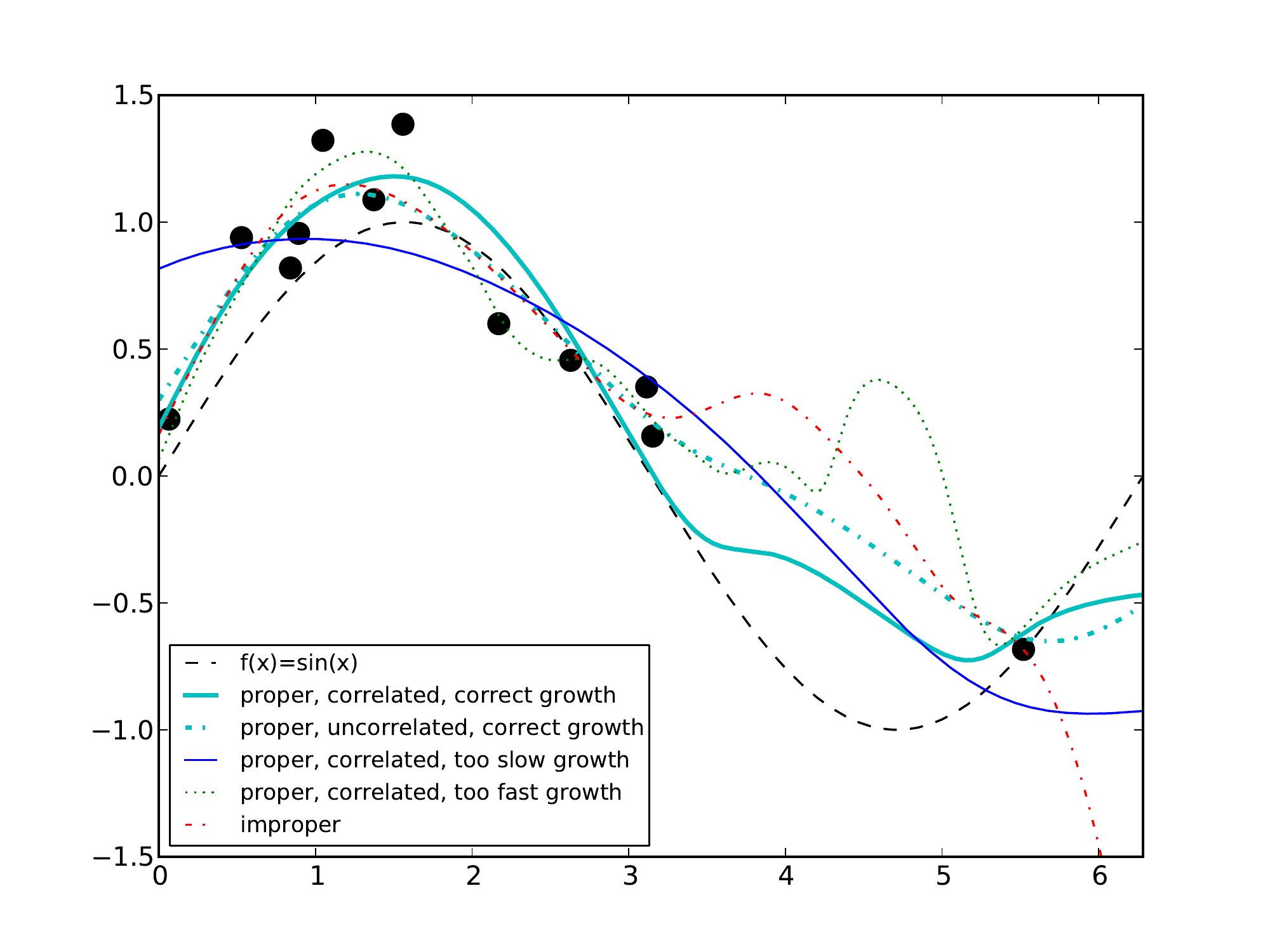}%
\end{centering}
\caption{\label{fig:sin_noerr_proper_improper}
(Colour online) 
Typical effect of priors for data points placed uniformly at random (black dots) 
from $f(x)=\sin(\omega x)$ (thin dashed black line),
with $\omega=1$ and $p=5$. Left: $N=6$, no value error. Right: $N=12$, medium-sized iid value error with $\sigma_\eps=1/4$.
In 100 such samples, the median $L_2$ distance between true and estimated function 
in the left[right] case was
0.068[0.24] for proper correlated $\phi$-priors growing as $\sigma_\phi\ui k\propto\omega^k$ with the correct frequency $\omega$ (thick solid cyan line),
0.15[0.29] for proper uncorrelated priors of the same size (thick dash-dotted cyan line),
0.22[0.48] for proper correlated priors with $\sigma_\phi\ui k\propto(\omega/2)^k$ (thin solid blue line),
0.25[1.7] for improper priors (thin dash-dotted red line),
and 0.56[0.55] for proper correlated priors with $\sigma_\phi\ui k\propto(2\omega)^k$ (thin dotted green line).
}\end{figure}


\subsection{Priors for oscillatory behaviour}

\paragraph{Correlation of derivatives.}
Assume that $d=1$ and $f(x)$ is of the form 
\begin{align}
    f(x) &= \int_0^{\infty}d\omega\,F(\omega)\sin(\omega x+t_\omega)
\end{align}
with a spectrum $F(\omega)$ and phases $t_\omega$ chosen uniformly at random.
Then the covariance of $f\ui k(x)$ and $f\ui\ell(x)$ is
\begin{align}\label{eqn:corr}
    \Sigma_\phi\ui{k,\ell} &= 
    \frac 1 2\int_0^{\infty}d\omega\,F(\omega)^2\omega^{k+\ell}\cdot
    \left\{\begin{array}{rl}
       1 & \text{if~} k-\ell = 0\mod 4\\
       0 & \text{if~} k-\ell \in\{1,3\}\mod 4\\
       -1 & \text{if~} k-\ell = 2\mod 4
    \end{array}\right.
\end{align}  
In other words, the $k$-th and $(k+4)$-th derivatives are fully correlated,
and the $k$-th and $(k+2)$-th derivatives are fully anti-correlated
(see also \citet{Gibson1992}).

If the spectrum of $f$ is known approximately, 
Eq.\,\ref{eqn:corr} the above is therefore a natural choice for the $\phi$-prior covariance structure.
For unknown spectrum, however, using a prior with strong correlation between derivatives
can result in worse results than using an uncorrelated or improper prior,
as can be seen in the example in Fig.\,\ref{fig:sin_noerr_proper_improper}.

\paragraph{Growth of derivative variance.}
If $F(\omega)$ is concentrated at a single frequency $\omega_0$, 
\ie, $f(x)=F_0\sin(\omega_0 x+t_0)$, the variance of $f\ui k(x)$ 
is growing exponentially: $\Sigma_\phi\ui{k,k} = F_0^2\omega_0^{2k}/2$.
For a broader but bounded spectrum, it is growing sub-exponentially.
\Eg, a uniform spectrum of the form $F(\omega)=F_0$ for $\omega\leq\omega_0$ and $F(\omega)=0$ for
$\omega>\omega_0$ gives $\Sigma_\phi\ui{k,k} = F_0^2\omega_0^{2k+1}/(4k+2)$.
For typical unbounded spectra, in contrast, the variance growth is super-exponential.
\Eg, for an exponentially decaying spectrum with $F(\omega)=F_0 e^{-a\omega}$ for some $a>0$, 
as is often assumed of chaotic dynamical systems, 
we get $\Sigma_\phi\ui{k,k} = F_0^2(2k-1)!k/(2a)^{2k+1}$. 
Similarly, for a Gaussian decaying spectrum with $F(\omega)=F_0 e^{-\omega^2/2v}$ for some $v>0$, 
we get $\Sigma_\phi\ui{k,k} = F_0^2v^{k+1/2}\Gamma(k+1/2)/4$. 
In case of a power-law decaying spectrum, the variance even diverges for large $k$.
\Eg, for $F(\omega)=F_0(1+a\omega)^{-b}$, 
we get $\Sigma_\phi\ui{k,k} = F_0^2\Gamma(2k+1)\Gamma(2b-2k-1)/2a^{2k+1}\Gamma(2b)$
for $k<b-1/2$ and $\Sigma_\phi\ui{k,k}=\infty$ for $k\geq b-1/2$,
justifying the use of an improper prior.

\subsection{Data-driven choice of remainder prior}

Without argument error, 
the {\em posterior predictive distribution} of $y\ui i$ 
from the Taylor model about $\xi=x_i$ has
\begin{align}
    \E(y\ui i) &= \mu_\phi\ui 0(\xi=x_i), &
    \Var(y\ui i) &= \tilde\Sigma_\phi\ui{0,0}(\xi=x_i) + \Sigma_\eps\ui{i,i}.
\end{align}
If the model is correct, one therefore expects that approximately
\begin{equation}\label{eqn:postpredglobal}
	\sum_i \frac{|y\ui i-\mu_\phi\ui 0(\xi=x_i)|^2}{\tilde\Sigma_\phi\ui{0,0}(\xi=x_i) + \Sigma_\eps\ui{i,i}} = N.
\end{equation}
Likewise, for any fixed $\xi$, and if $P_\phi=0$ and $\mu_\psi=0$, 
the posterior predictive distribution of $W^{1/2}\bm y$
from the Taylor model about $\xi$ has
\begin{align}
    \E(W^{1/2}\bm y) &= W^{1/2}X\tilde{\bm\mu}_\phi,\\
    \Cov(W^{1/2}\bm y) &= W^{1/2}X\tilde P_\phi^{-1}X'W^{1/2} + W^{1/2}(\Sigma_r+\Sigma_\eps)W^{1/2}\\
    	&= W^{1/2}X(X'WX)^{-1}X'W^{1/2} + I.\nonumber
\end{align}
If the model is correct, one therefore expects that approximately
\begin{equation}\label{eqn:postpredlocal}
    ||W^{1/2}(X\tilde{\bm\mu}_\phi-\bm y)||_2^2 = \trace\{W^{1/2}X(X'WX)^{-1}X'W^{1/2} + I\} = m + N.
\end{equation}
Hence, assuming $\Sigma_\psi = vS$ with a known matrix $S$ and unknown $v>0$, 
one could either use the same $v$ for all $\xi$ and choose it so that the ``global'' Eq.\,\ref{eqn:postpredglobal} is fulfilled,
which is similar to the suggested parameter choice in \citet{Wang2010}, 
or use a different $v$ for each $\xi$ and choose it so that the ``local'' Eq.\,\ref{eqn:postpredlocal} is fulfilled. 

\subsection{Interpolation with noninformative priors}
If $\Sigma_\eps=0$ and $\Sigma_\psi=v I$, $\tilde{\bm\mu}_\phi$ does not depend on $v$,
and Eq.\,\ref{eqn:postpredlocal} becomes
\begin{align}
	v&=||D^{1/2}(X\tilde{\bm\mu}_\phi-\bm y)||_2^2/(m+N), & D\ui{i,j} &= \delta_{ij}/\textstyle\sum_{\alpha,|\alpha|=p}(X\ui{i,\alpha})^2
\end{align}
(see Eq.\,\ref{eqn:postr}).
This value is also identical to the posterior mode of $v$
when $v$ is treated as a scale hyperparameter with a noninformative Jeffreys prior of $\rho(v)\propto 1/v$.
Hence the \motabar\ interpolant with noninformative priors is
\begin{align}
	\bm{\tilde\mu}_\phi &= (X'DX)^{-1}X'D\bm y, &
    \tilde\Sigma_\phi &= \frac{||D^{1/2}(X\tilde{\bm\mu}_\phi-\bm y)||_2^2}{m+N}(X'DX)^{-1}.
\end{align}

%
%

\python{
# mathematica:

k = 4
p = 2k + 1
X = Table[i^a / a!,{i,-k,k},{a,0,p-1}]
X[[k+1,1]] = 1
S = DiagonalMatrix[Table[1+0*i^(2p),{i,-k,k}]]
W = PseudoInverse[S]
W // MatrixForm
A = PseudoInverse[Transpose[X].W.X].Transpose[X].W
(p-2)! A // MatrixForm
}

\section{Example: reconstruction of Lorenz attractor from noisy observations}

\begin{figure}\begin{centering}%
(a)\includegraphics[width=0.45\textwidth]{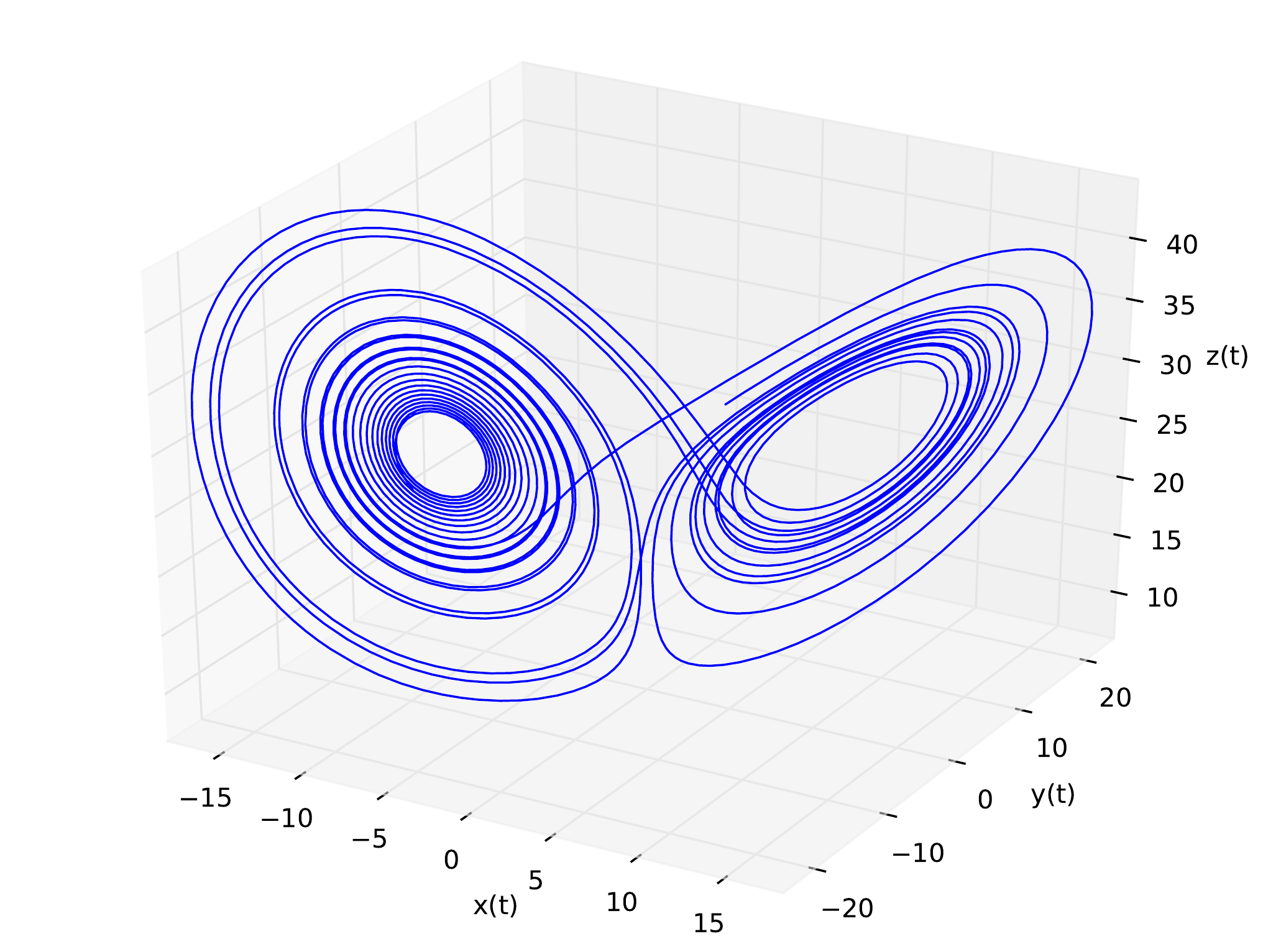}%
(b)\includegraphics[width=0.45\textwidth]{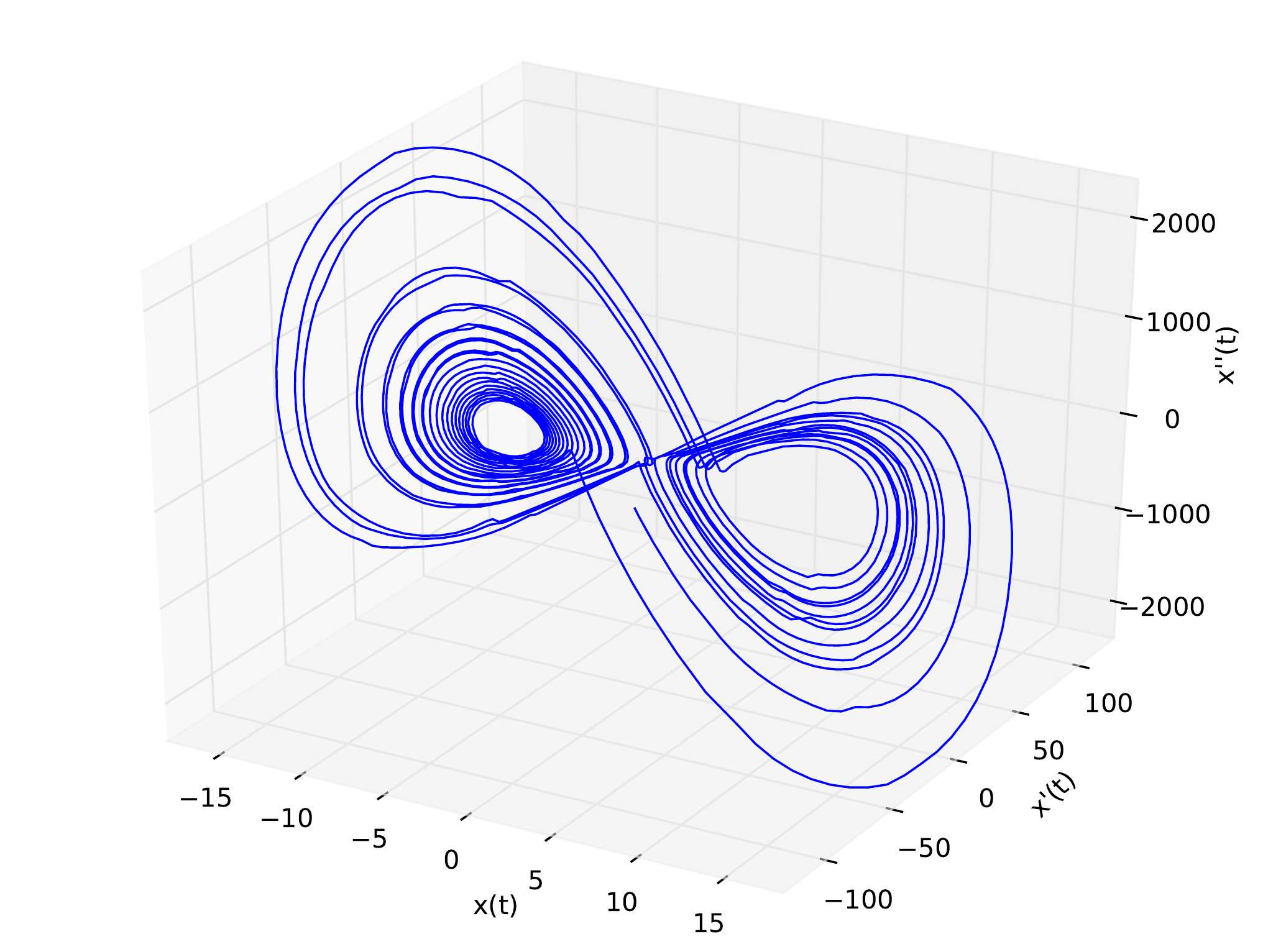}\\%
(c)\includegraphics[width=0.45\textwidth]{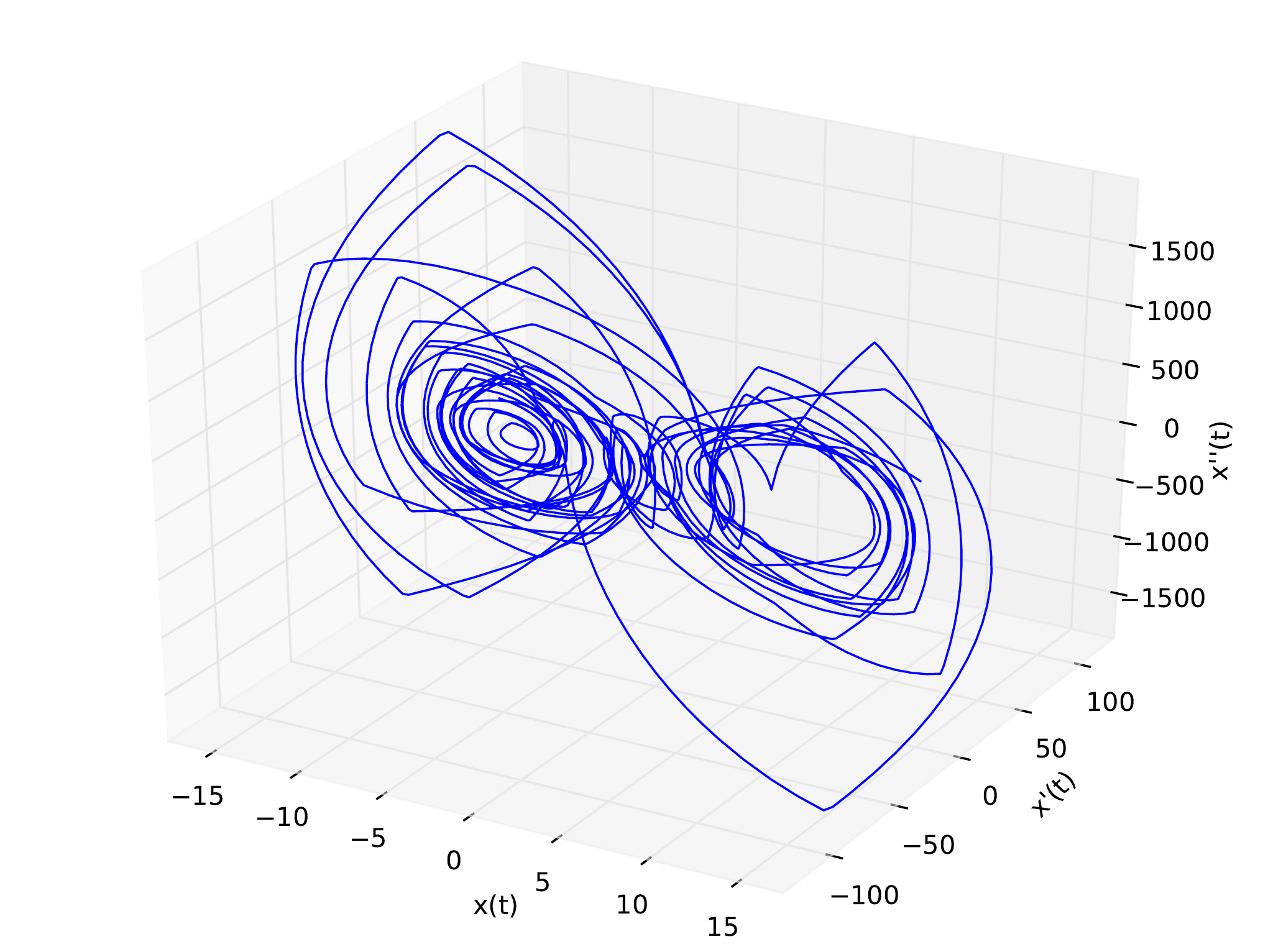}%
(d)\includegraphics[width=0.45\textwidth]{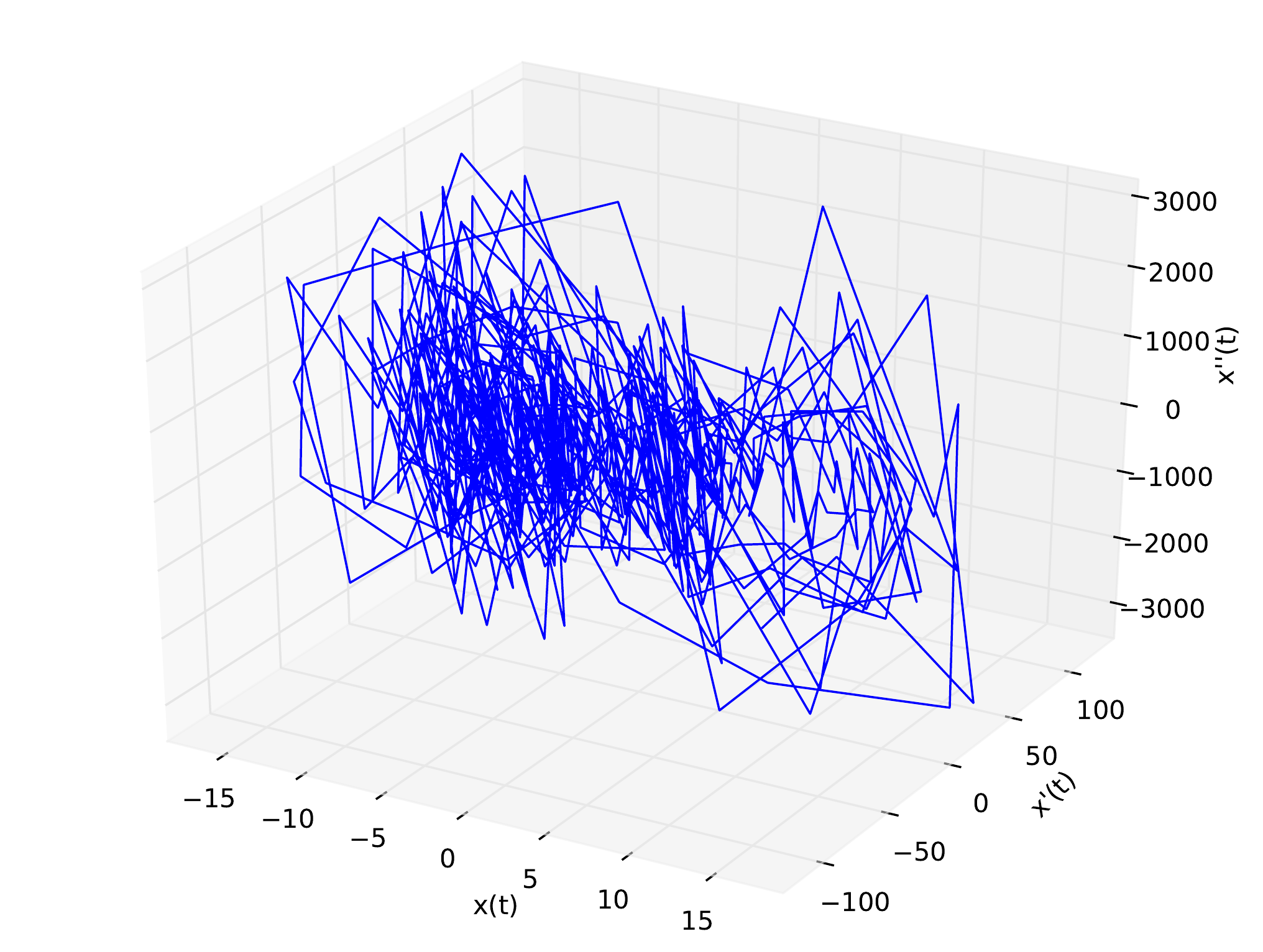}\\%
(e)\includegraphics[width=0.45\textwidth]{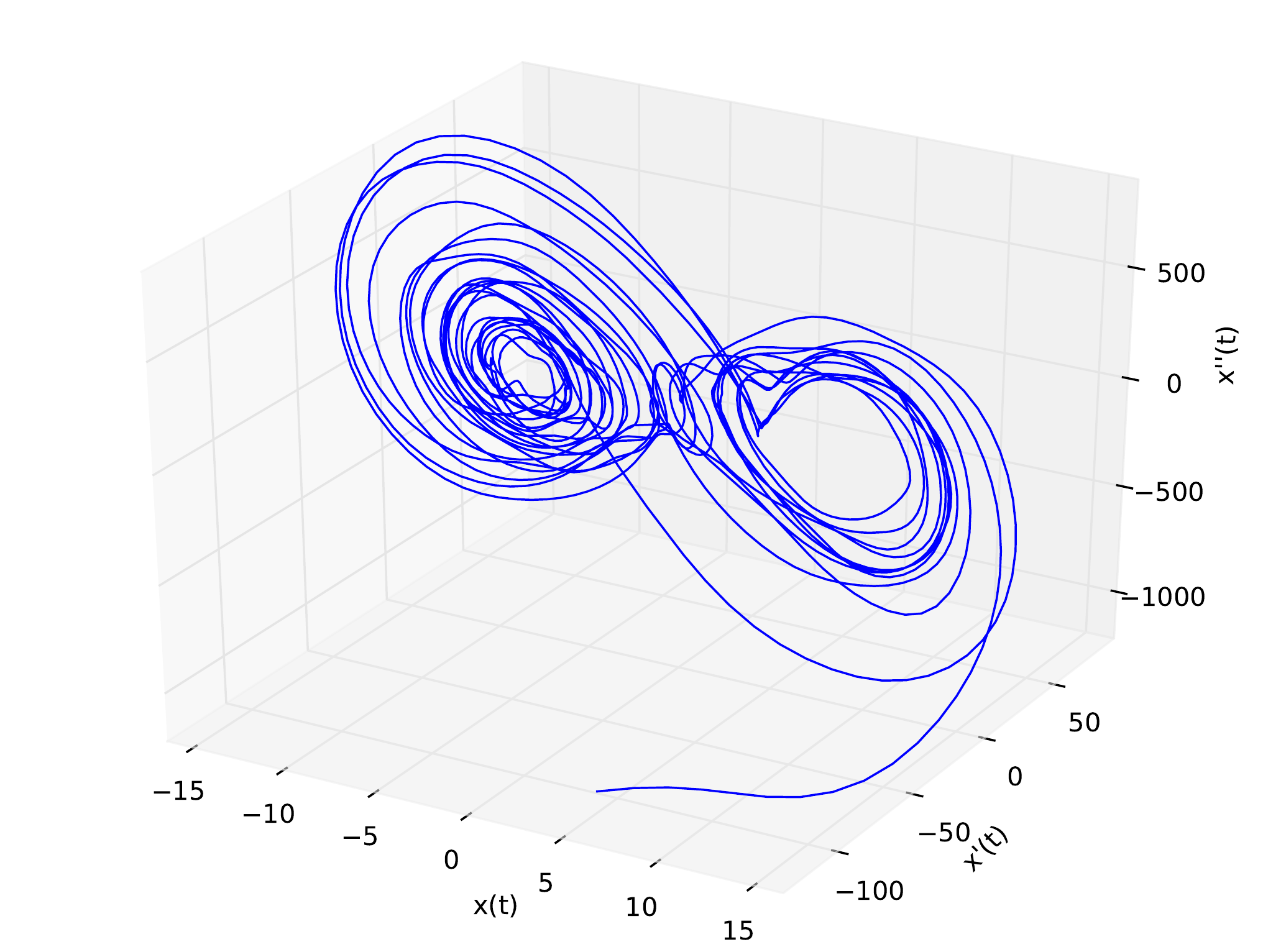}%
(f)\includegraphics[width=0.45\textwidth]{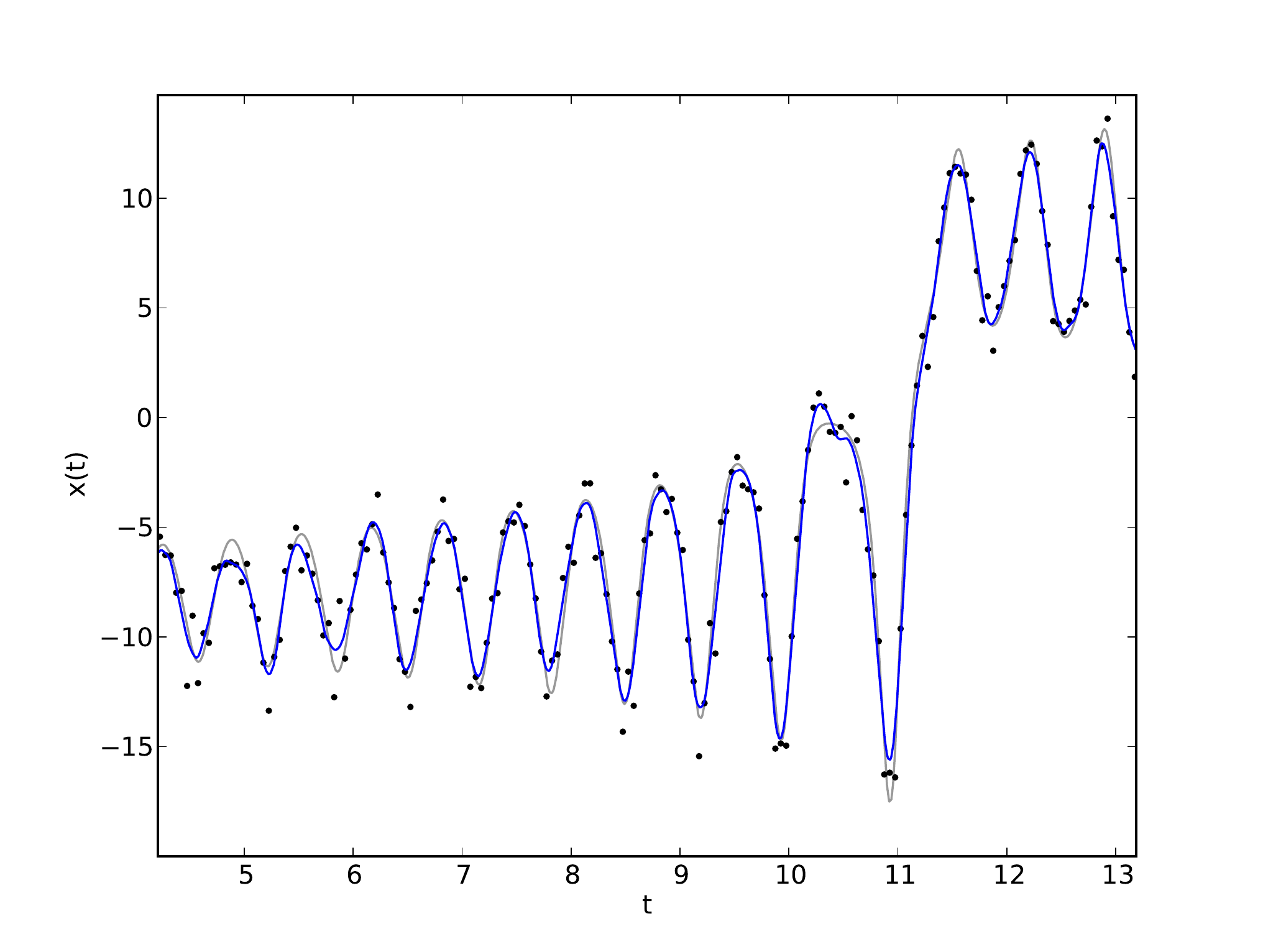}%
\end{centering}
\caption{\label{fig:lorenz}
(Colour online) 
Reconstruction of the Lorenz attractor from noisy measurements of its $X$ component.
(a) Simulated true phase space trajectory.
(b) Derivatives of $X$ from numerical differentiation from simulated true $X$ data.
(c) Derivatives from natural cubic smoothing spline for noisy measurements of $X$.
(d) Numerical differentiation from noisy measurements.
(e) Derivatives from \motabar\ with $p=4$ and an improper $\phi$-prior.
(f) Detail of \motabar\ estimated $X$ (thick blue line) compared to simulated true $X$ (thin gray line),
and noisy measurements (black dots).
}\end{figure}

To demonstrate the performance of \motabar\ for complex data,
we simulated a trajectory $(X,Y,Z)(t)$ of the standard chaotic Lorenz system
given by the ODEs 
$\dot X=10(Y-X)$, $\dot Y=-XZ+28X-Y$, $\dot Z=XY-10Z/3$,
where $X$ basically oscillates between $\approx\pm 15$ with a frequency of $\approx\omega=8$
(see Fig.\,\ref{fig:lorenz},a).
We then generated a sample of $N=500$ noisy observations 
$X(t_i)+\eps(t)$, with iid errors $\eps(t)\sim N(0,1)$ (which corresponds to $\approx 1\%$ of noise),
for time-points $t_i$ that were regularly spaced at a distance $\Delta t=0.05$
(which corresponds to $\approx 15$ data points per oscillation, see the black dots in Fig.\,\ref{fig:lorenz},f).
From the sample, we reconstructed the original trajectory (Fig.\,\ref{fig:lorenz},a) 
up to a diffeomorphism,
following the approach of \citet{Packard1980} 
by estimating the time evolution of the derivatives $(X,\dot X,\ddot X)$ (Fig.\,\ref{fig:lorenz},b),
using different estimation methods.
For \motabar\ estimation, we used $p=4$, an improper $\phi$-prior,
and a $\psi$-prior whose variance $\sigma^2_\psi = 15\omega^p I$ was chosen to fit the approximate range and frequency of the oscillation.
Fig.\,\ref{fig:lorenz}(e) shows that the \motabar\ approach reproduces the shape of the trajectory much better than
either spline smoothing (c) or numerical differentiation (d),
the latter being basically equivalent to an alternative phase space reconstruction method, 
the often used ``method of delays''.
Note that there are other, more sophisticated state space reconstruction methods
based on PCA or discrete Legendre polynomials, 
which deal better with noise \citep{Gibson1992}
and which will be compare to \motabar\ in a separate paper.



\section{Conclusion}

\paragraph{Outlook: alternative local approximations.}
The simple form of the \motabar\ estimator is due to our assumption of Gaussian value measurement errors 
and the fact that Taylor polynomials are linear in their coefficients.
Alternatively, one might approximate $f$ by other functions 
that can be parameterised by the low-order derivatives of $f$ at $\xi$.
For $d=1$ and if it is known that $f$ displays oscillatory behaviour, one such approximation could be
\begin{align}
    f(x) = a + b\sin(\omega(x-\xi)) + c\cos(\omega(x-\xi)) + r(x-\xi)
\end{align}
with a remainder function $r$ with $r(0)=r'(0)=r''(0)=r'''(0)=0$.
Because then 
$\omega=\sqrt{-f'''(\xi)/f'(\xi)}$,
$c=f'(\xi)f''(\xi)/f'''(\xi)$,
$a=f(\xi)-f'(\xi)f''(\xi)/f'''(\xi)$,
and
$b=f'(\xi)\sqrt{-f'(\xi)/f'''(\xi)}$,
the four derivatives can be estimated from the model if a plausible prior for $r(x-\xi)$ is used,
although the estimate will not be a linear function in $\bm y$ 
since the above approximation is not linear in the coefficients.  
If $f$ is known to be periodic with frequency $\omega$,
it might seem that one could also use partial sums of the corresponding Fourier series instead,
which are linear in their coefficients,
but they are not parameterizable by a finite number of derivatives of $f$ at $\xi$. 


\paragraph{Software.}
An open-source software package implementing \motabar\ for use with the python programming language 
is under development and will be made available at 
\url{http://www.pik-potsdam.de/members/heitzig/motabar}.

\section*{Acknowledgements}
This work was supported by the German Federal Ministry for Education and Research (BMBF) 
via the Potsdam Research Cluster for Georisk Analysis, Environmental Change and Sustainability (PROGRESS).
The author thanks Forest W.~Simmons, Kira Rehfeld, Norbert Marwan, Bedartha Goswami, and J\"urgen Kurths for fruitful discussions.


\end{document}